\documentclass[universe,review,accept,moreauthors,pdftex]{Definitions/mdpi} 

\firstpage{1} 
\pubvolume{7}
\issuenum{5}
\articlenumber{118}
\pubyear{2021}
\copyrightyear{2021}
\externaleditor{Academic Editor: Roberto Mignani}
\datereceived{25 March 2021} 
\dateaccepted{20 April 2021} 
\datepublished{22 April 2021} 
\hreflink{https://doi.org/10.3390/universe \linebreak 7050118} 

\Title{``In-System'' Fission-Events: An Insight into Puzzles of 
Exoplanets and Stars?
}

\TitleCitation{``In-System'' Fission-Events: An Insight into Puzzles of 
Exoplanets and Stars?}

\Author{Elizabeth P. Tito $^{1,}$*\orcidA~and Vadim I. Pavlov $^{2,}$*\orcidB} 

\AuthorNames{Elizabeth P. Tito and Vadim I. Pavlov}
\AuthorCitation{Tito, E.P.; Pavlov, V.I.}
\address{%
$^{1}$ \quad Scientific Advisory Group, Pasadena, CA 91125, USA\\
$^{2}$ \quad Facult\'{e} des Sciences et Technologies, Universit\'{e} de Lille, F-59000 Lille, 
France}

\corres{Correspondence: eptito@gmail.com (E.P.T.); Vadim.Pavlov@univ-lille.fr (V.I.P.)}

\label{Abstract}
\abstract{
In expansion of our recent proposal 
that the solar system's evolution occurred in two stages---during the first stage, the gaseous giants formed (via disk instability), and, during the second stage (caused by an encounter with a particular {\em stellar-object} leading to ``in-system'' fission-driven nucleogenesis), the terrestrial planets formed (via accretion)---we emphasize here that the mechanism of {\em formation} of such stellar-objects is generally universal and therefore {\em encounters} of such objects with stellar-systems may have occurred elsewhere across galaxies.  If so, their {\em aftereffects} 
may perhaps be observed as puzzling features in the spectra of individual stars (such as idiosyncratic chemical enrichments) and/or in the structures of exoplanetary systems (such as unusually high planet densities or short orbital periods). This paper reviews and reinterprets astronomical data within the ``fission-events framework.''  Classification of stellar systems as ``pristine'' or ``impacted'' is offered. 
}


\begin{document}

\newcommand{\araa}{Annu.~Rev.~Astron.~Astrophys. }
\newcommand{\aaa}{Astron.~and~Astrophys. }
\newcommand{\aap}{Astrophys.~Astron. }
\newcommand{\aj}{Astronom.~J } 
\newcommand{\apj}{Astrophys.~J}
\newcommand{\aplett}{Astrophys.~Lett.} 
\newcommand{\aphjl}{Astrophys.~J.~Lett. }
\newcommand{\apjl}{ApJ Lett. }
\newcommand{\apspr}{Astrophys.~Space~Phys.~Res. }%
\newcommand{\apss}{Astrophys.~Space~Sci. }%
\newcommand{\gal}{Galaxies }
\newcommand{\gca}{Geochimica et Cosmochimica Acta} 
\newcommand{\icarus}{Icarus }
\newcommand{\jcap}{J.~Cosmology and Astroparticle Phys.} 
\newcommand{\mnras}{Mon.~Not.~R.~Astron.~Soc. }
\newcommand{\nar}{New~Astron.~Rev.} 
\newcommand{\nat}{Nature }
\newcommand{\nphysa}{Nucl.~Phys.~A } 
\newcommand{\prl}{Phys.~Rev.~Lett. }
\newcommand{\prc}{Phys.~Rev.~C }
\newcommand{\prd}{Phys.~Rev.~D }
\newcommand{\pasp}{PASP}
\newcommand{\sovast}{Soviet Ast.-AJ }
\newcommand{\areps}{Annu.~Rev.~Earth~and~Planet.~Sci.}
\newcommand{\arnps}{Annu.~Rev.~Nucl.~Part.~Sci.}
\newcommand{\epsl}{Earth~and~Planet.~Sci.~Lett.}
\newcommand{\jgr}{J.~Geophys.~Res.}
\newcommand{\jpgnpp}{J.~Phys.~G:~Nucl.~Part.~Phys.}
\newcommand{\mps}{Meteoritics~and~Planet.~Sci.}
\newcommand{\pl}{Phys.~Lett.}
\newcommand{\rmp}{Rev.~Modern~Phys.}


%
%

\section{Introduction}
 
As facilities and techniques for astronomical observations and analyses continue to expand and gain in resolution power, their results provide increasingly detailed information about stellar systems, in particular, about the chemical compositions of stellar atmospheres  and structures of exoplanets. 
A number of puzzles have been discovered. 
For example, for some stellar systems, enrichment with certain chemical elements appears to unexplainably deviate from expectations of galactic nucleosynthesis models \cite{Jacobson_2014}. 
Another set of puzzles concerns exoplanetary systems 
\cite{Marov_2020,Borucki_2017,Raymond_2014}.     
When compared to the solar system, exoplanetary systems seem different.  
Very compact planetary systems with planets orbiting close to their host star are frequent. 
The characteristics of many exoplanets are also unlike those found in the solar system. 

Historically, a simple paradigm 
\cite{Cameron_1962} 
was invoked to explain the evolution and structure of planetary systems.  
This paradigm assumed that each planetary system would form from a co-rotating disc of gas and dust that resulted from a collapsing portion of a giant molecular cloud. 
Rocky planets would form only near the star because the temperature was too high for non-refractory materials to condense. 
At farther distances, the temperatures were cool enough to condense volatile materials. 
Accreted planet embryos would quickly sweep up enough material to become so massive that the hydrogen and helium in the protoplanetary disk would be captured, producing ice- and gas-giant planets. 
Although some of the assumptions in this paradigm are consistent with observations of exo-systems, 
revision and extension of this paradigm are active areas of research 
(see \cite{Marov_2020,Borucki_2017,Marov_2018}, and references therein). 

In our recent publication \cite{Tito_2020}, we advanced a hypothesis (of fission-driven nucleogenesis in the solar system) offering an ``expanded'' paradigm for how the solar system formed and evolved. 
In brief, we proposed that the solar system was ``impacted'' at one point of its history, and, consequently, its lifetime may be divided into two stages: pre-impact and post-impact. 
During the first stage, which started {\em earlier} than $\sim$4.56~Gyrs ago, 
{\em only} the Sun and the gaseous giants formed 
(via gravitational disk instability from the nebula not-yet-enriched with exotic and ``heavy'' nuclei). 
The structure of the first-stage solar system then resembled one type of exoplanetary system:  
those which are comprised of gaseous giants only, located at substantial distances of order of $10^0$--$10^2$~AU from the host star. 
We then proposed that, $\sim$4.56~Gyrs ago, a fission-driven nucleogenetic event 
(the physics of which was elaborated in Ref. \cite{Tito_2020}) 
occurred in the inner part of the solar system. 
During the second stage, 
the rocky inner planets formed (via accretion) from the debris from the event, 
and the Sun and the gaseous giants captured some of the debris, thus becoming enriched with exotic nuclei.  
We suggested that the event was caused by the arrival and 
explosion (in the inner part of the solar system) 
of a stellar-object born long time ago 
in a distant galactic cataclysm, perhaps involving the central supermassive black hole 
Sagittarius~$A^*$ (Sgr~$A^*$).   
Specifically, we suggested that such object was a giant droplet-shaped ``clump'' of super-dense nuclear-matter (quasi-stable and transitionable into nuclear-fog, 
thus leading to fission-driven nucleogenesis when sufficiently perturbed). 
 
In this review-paper, we emphasize that the mechanism of formation of such stellar-objects 
is generally {\em universal} and therefore encounters of such objects with stellar-systems may have occurred elsewhere across galaxies.  
If so, we may have already observed some of their aftereffects in the ``impacted'' stellar systems, and we may be able to detect more ``impact'' signatures in future observations once we know what to look for.  

The goal of this paper is to identify and discuss such observable signatures. 

Obviously, the outcomes of such encounters would not necessarily have been identical to what happened in the solar system.  
In some systems, the exotic traveller may have been captured by the central star---possibly revealing itself via exotic features in the stellar atmosphere's spectrum. 
In some systems, it may have exploded very close to the star---this could explain, for example, why the subsequently-formed planets have unusually high-densities and/or appear very close to the host.    
Naturally, many stellar systems most likely have not yet encountered such objects (and may never encounter one), remaining ``pristine'' in their composition and structure. 
In other words, two broad categories of stellar systems may in fact exist: ``{\em pristine}'' and ``{\em impacted}''. 
If so, the entirety of available data---on stellar spectra, exoplanetary characteristics, sudden ``changes'' (in spectra or brightness), interstellar or stellar ``explosions'', and so on---may contain new insights or offer new interpretations if re-examined in the framework of this hypothesis. 
To facilitate the start of such re-examination, we first briefly summarize the key aspects and mechanisms of the ``fission-events framework'' (Section~\ref{s:2}).  Then, in the context of this framework, we discuss certain data on stellar chemical compositions (Section~\ref{s:3}) and exoplanetary systems (Section~\ref{s:4}). The final section summarizes key distinctions between the conventional and fission-events frameworks, and the features of ``candidates'' for being ``impacted'' systems. 

\section{The ``Fission-Events Framework''} 
\label{s:2}

It is well known that stellar ``fragments''---pieces of stellar objects---can be formed and ejected with substantial velocities in a number of stellar cataclysms. 
For example, during the rotating core collapse, one or more self-gravitating lumps of nuclear-like matter can form in close orbit around the central nascent neutron star \cite{Imshennik_1998}.
The unstable (in the phase-transition and nuclear-reaction sense) member of such transitory multi-fragment system ultimately explodes,
giving the surviving member a substantial kick velocity---as fast as $10^3$--$10^4$~km/s \cite{Colpi_2002}. 
Massive black holes can tear apart even such compact super-dense objects as neutron stars, and then catapult some of the pieces \cite{Rees_1990,Tito_2018b}. 
It is also well known that, at the center of our galaxy 
($\sim$7.92~kpc away \cite{Hirota_2020}), lies the supermassive black hole Sgr~$A^*$ 
with mass $\sim$4.3 $\times 10^6 M_{\odot}$ \cite{Gillessen_2017}. 
Its crushing power is sufficient to generate numerous ``clumps'' of super-dense nuclear-matter and send them off across the galaxy. 
Hyper-velocity $\sim$1700~km/s was recently observed even for a main sequence star kicked by Sgr~$A^*$ 4.8~Myr ago with the implied velocity $\sim$1800~km/s  \cite{Koposov_2019}. 
 Furthermore, an entire tidal disruption event---the closest to date at 66~Mpc---was recently observed, 
 in which the host-galaxy's black hole with mass $\sim$$10^6 M_{\odot}$ partially disrupted a 
 $\sim$1 $M_{\odot}$ star.  About 75 percent of the star was stripped during the encounter.    
The early spectra showed emission lines consistent with electron scattering in an expanding medium with $v \sim$3000--10000~km/s \cite{Nicholl_2020}. 
 
Although trajectories of small objects cannot be traced at great distances, 
in recent years, two small interstellar visitors have been detected in the solar system:  
object 'Oumuamua, officially named 1I/2017 U1 (moving at interstellar speed, i.e., velocity at infinity, of $\sim$26~km/s and passing at 0.16~AU from Earth and 0.25~AU from the Sun) and comet 2I/Borisov (moving at $\sim$42~km/s and passing the Sun at $\sim$2~AU).  
Their eccentricities, 1.20 for ‘Oumuamua and 3.7 for 2I/Borisov, indicate that they have never been gravitationally bound to the Sun.  
Their origins and compositions (certainly not composed of any nuclear-matter) 
are currently being explored; see, for example, \cite{Zhang_2020,Bodewits_2020}. 
These detections, however, illustrate that small objects of various origins indeed traverse the interstellar space, even in the solar system's neighborhood of the galaxy. 

For the stellar ``in-system'' fission-events to occur, the black-hole catapulted ``objects'' have to (1) be composed of {\em super-dense nuclear-matter}, (2) remain internally stable during the journey, but (3) lose their stability once perturbed 
(see Appendix~\ref{appa}, or Ref. \cite{Tito_2020} for details and references). 
Thus, their nuclear-matter has to be {\em quasi-stable} at the moments of encounters and become unstable as the result of the experienced perturbations, thus starting the ``in-system'' fission-driven nucleogenetic cascades and nuclear explosion. 
This process occurs when the nuclear-matter enters its state of {\em nuclear-fog}, which, unlike the familiar water-fog, is in fact explosive (see Appendix~\ref{appa}).    

To summarize briefly what is explained in detail in Ref. \cite{Tito_2020}:   
At first---and for a very long time---the torn-away ``clumps''  of nuclear-matter travel across the galaxy  
remaining structurally-stable. 
(It has been theoretically demonstrated that objects composed of dense nuclear matter but smaller, even significantly smaller, than conventional neutron stars, or perhaps other exotic stars—can indeed exist in a drop-like form, staying as dense as a nucleus, and remaining structurally stable \cite{Tito_2018a}.)
However, gradually, the clumps  cool down. 
When the clump cools down ``enough,'' its thermodynamical state reaches the boundary of the two-phase zone of instability,  
the matter becomes thermodynamically unstable, enters the state of nuclear-fog, 
and any further perturbation sets off 
nuclear-fragmentation and nuclear-fission of its separated mega-nuclei, 
again and again, which combine with the full set of various captures and decays possible in the environment that is neutron-rich (see Appendix~\ref{appa}).   
The result is a nuclear (not thermonuclear) explosion producing nucleogenetic (fission-driven) cascades. 

Overall, four main types of scenarios may be envisioned: 
\begin{itemize}
\item[(1)] a fission-event may occur in the ``interstellar space''---perhaps triggered by perturbations due to 
encountered variations in interstellar medium/fields or 
propagating ejecta/shock wave from some other stellar cataclysm---thus enriching the medium (the molecular cloud and subsequent protonebulas) in the vicinity; 
\item[(2)] a fission-event may occur in a protonebula, thus enriching the nebula (producing either homogeneous or heterogeneous distribution of nuclei) and perhaps serving as the trigger for the nebula's collapse (as considered for the solar system, suggesting a supernova as the trigger); 
\item[(3)] a fission-event may occur in a protodisk, injecting new nuclei and abruptly changing disk properties; 
\item[(4)] a fission-event may occur within an already-formed stellar system,  
possibly impacting the host--star (its atmosphere and/or interior) and/or the orbital structure and compositions of the planets. 
\end{itemize}

Thus, it is possible that certain peculiar stellar spectra (Section~\ref{s:3}) and 
planetary characteristics (Section~\ref{s:4}) 
of individual exosystems 
are in fact indications of occurrences of fission-events during the lifetimes of the systems.  
(Appendix~\ref{appb} presents a brief discussion on the likelihood of fission-events within exosystems.)

\section{Chemical Compositions of Stars}
\label{s:3}

Generally speaking, chemical enrichment of individual stars is a product of successive cycles of star formation and evolution in the galaxies and specific neighborhoods. 
As well known---see, for example, Ref. \cite{Jacobson_2014}---the first stars in the universe (so-called Population III stars for historical reasons) formed from clouds of {\rm H} and {\rm He} (with maybe 
some {\rm Li}). 
Population III stars are thought to have been quite massive (10 to 100 $M_\odot$), and therefore expected to have exploded as supernovae a few million years after they formed, polluting interstellar gas with the products of {\em nucleosynthesis} in their interiors and in their supernovae. 
The subsequent generations (called Population II) inherited the chemical imprint of the first generation, and then further enriched the interstellar medium with products of their own nucleosynthesis in the final 
stages of their evolution (supernovae, or AGB stars). 
The early Population II stars are called “metal-poor” stars, to indicate the relative paucity of the products of stellar nucleosynthesis in their atmospheres, compared to that of the Sun, which almost always serves as the reference.    
{\rm Fe} is typically used as a proxy for metallicity because the large number of {\rm Fe} absorption lines present in the optical wavelength regime makes it straightforward to measure. 
Notation [A/B] describes the relative abundances of two elements in a star compared to that in the Sun: [A/B] = log$_{10}(N_A/N_B) - $log$_{10}(N_A/N_B)_\odot$. 
A star with {\rm [Fe/H]} = $-2$, for example, contains a factor 100 fewer {\rm Fe} nuclei 
by number than the Sun. 
The associated unit is a logarithmic unit: dex (contraction of ``decimal exponent''). 

Despite the significant progress in understanding of the galactic nuclei-enrichment processes, 
there remain some unresolved puzzles. 
For example, as stated in \mbox{References \cite{Jacobson_2014,Hinkel_2014,Venn_2004}}:   

\begin{quote}
\begin{enumerate}
\item  “Actinide Boost”:  
About a quarter of strongly $r$-process enhanced stars shows {\rm Th} abundances that are higher than expected compared to other stable $r$-process elemental abundances and the scaled solar $r$-process pattern. 
This results in {\em negative} stellar ages when using the {\rm Th/Eu} chronometer. 
«One explanation may be that these stars show the $r$-process pattern of two $r$-process events that occurred at different times---one just prior to the star’s formation and one at a later time in the vicinity of the star»  \cite{Jacobson_2014}. 
\item  «[A]ll the strongly $r$-process enhanced metal-poor stars found so far exhibit a narrow range in {\rm [Fe/H]} of 0.3--0.4~dex. If the $r$-process is universal, why do these stars appear at a certain “chemical time”, as put by \cite{Hansen_2011}? Some propose this signals the start of a new process at work in the chemical evolution of the universe (e.g., \cite{Francois_2007}), or else that these stars only form from a very special type of supernova in which the neutron-capture elements are released via jets, unlike the other elements \cite{Hansen_2011}»  \cite{Jacobson_2014}. 
\item	«[N]o stars with ${\rm [Fe/H]} < -3.5$ have yet been discovered that display any known or characteristic neutron-capture abundance pattern. This raises the question of when exactly the very first neutron-capture events took place in the early universe and whether massive Population III stars produced neutron-capture material, and, if so, in what quantities» \cite{Jacobson_2014}.  
\item «[A]bundances of neutron-capture elements with 40 < Z < 56, i.e., those between the first and second peak, signal that yet other, unidentified \dots 
processes may have been at work in the early universe. In their analysis of silver and palladium in metal-poor stars, \citet{Hansen_2012} found that the abundance ratios of {\rm Pd} and {\rm Ag} (e.g., {\rm [Ag/Fe], [Ag/Eu], [Ag/Ba]}) did not match the patterns expected if they were produced by the main $r$, the weak $r$, or any $s$-process channel»  \cite{Jacobson_2014}.
\item «{\rm [Fe/H]}-rich group near the midplane is deficient in {\rm Mg, Si, S, Ca, Sc II, Cr II}, and {\rm Ni} as compared to stars farther from the plane» \cite{Hinkel_2014}. 
\item Observations of the ratios of {\rm [Y/Fe], [Ba/Fe], [La/Fe]}, and {\rm [Eu/Fe]} in the stars in the Milky Way dwarf spheroidal satellite galaxies (dSphs) and the stars in the Galaxy indicated that 
«{\rm [Y/Fe]} is significantly lower/offset in the dSph stars than in the Galaxy. This includes roughly half of the dSph stars, and suggests the $r$- and $s$-process enrichment of this element differs between the galaxies \dots 
This result suggests that  the site of $r$-processed {\rm Y} must differ from that of $r$-processed {\rm Ba, La}, and {\rm Eu}; is there a weak $r$-process site? In addition, the source that produces {\rm Y} in the metal-poor Galactic stars must be absent in the dSphs or it must have a different time lag relative to the {\rm Ba, La}, and {\rm Eu} enrichments  \dots 
[N]o population of stars in the Galaxy is representative of stars in the low mass dwarfs» 
\cite{Venn_2004}.  
\end{enumerate}
\end{quote}

The proposed fission-event framework offers an {\em additional} nuclei-enrichment mechanism to the existing inventory of considered mechanisms.  
The nuclei-production signature of {\em nuclear-fission} (from higher nucleon numbers $A$ to lower $A$) cascading from extremely large nucleon numbers ($ln A \gg 1$) necessarily differs from the signature of {\em nucleosynthesis} (from lower $A$ to higher $A$). 
The signature of {\em fission-cascade} also differs from the so-called {\em fission-(re)cycling} sometimes 
 incorporated in models---mainly in the models of mergers of neutron stars with other neutron stars or with black holes---which occurs once the {\em upward} nucleosynthesis forms fissioning nuclei, such as actinides and super-heavy nuclei immediately above actinides (with model-limits at $A \sim$320 as no further data exists; see, for example, \cite{Shibagaki_2016}).  
In {\em fission-cascade}, the {\em downward} fission (multi-fragmentation) of a giant ``clump''  of super-dense nuclear-matter differs greatly from such models because: 
(1)~it starts not with $A \leq 320$ (as conventionally modeled) but with $ln A \gg 1$; 
(2)~it manifests itself in nuclegenetic cascades involving very short-lived nuclei (never considered in conventional models); 
and  
(3)~it produces distribution of final nuclei which is unpredictable (rather than with model-specified probabilities). 
Ultimately, {these cascades produce all types of nuclei: ``heavy'' nuclei (super-heavy, actinides, post-{\rm Fe}); $r$-, $s$-, and $p$-nuclei; stable and short-lived nuclei}; and so on.  
Modeling such cascades, however, requires further advancements in nuclear physics, both experimental and theoretical. 
For full discussion, see Ref. \cite{Tito_2020}.  

\begin{figure}[H]
 \includegraphics[width=0.93\columnwidth]{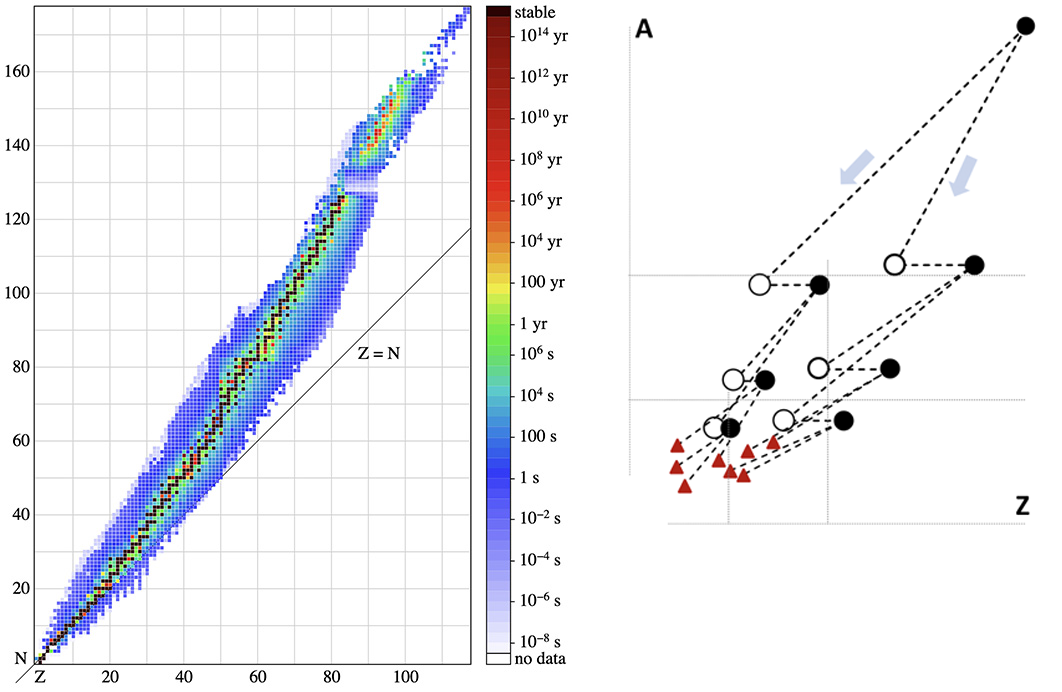}  
\caption
[Fission-Events Framework: Nucleogenetic Cascades (Isotope Production Dispersion)]
{ 
{\bf Left Panel}:  Half-lives of isotopes (Z is a number of protons and N is a number of neutrons in a nucleus). 
Data from National Nuclear Data Center (NuDat2 database 6/1/2012) \cite{Wiki_2012}. 
{\bf Right Panel}: Dispersion in nucleogenetic cascades in ZA-plane (where A = Z+N). 
For visual simplicity, only fission and $\beta$-decays are depicted.
}
%
\label{Fig:1} 
\end{figure}

Figure~\ref{Fig:1} is provided to help intuitively appreciate why and how all types of nuclei may be produced.  

The left panel shows the half-lives of the known isotopes. 
Thus far, only elements up to $Z = 118$ have been explored \cite{Oganessian_2015,Oganessian_2017,Giuliani_2019}, and half-lives of only a limited number of isotopes of each known element have been estimated (shown as colored cells).  
However, the vast white space in the chart represents the yet-unexplored very-short-lived nuclei---they cannot be ignored in the top-down fission of mega-nuclei. 
The cascades starting with gigantic nucleon numbers ($ln A \gg 1$) would go not only along the ``colored'' zone but also through the ``white'' zones---on both sides of the colored zone and above the depicted domain, extending to much higher $A$. 

The right panel provides a simple illustration of how a cascade 
(which includes only asymmetric fission and $\beta$-decays for visual simplicity) 
creates a dispersion of nuclei. 
Even after just three steps, it is apparent that the produced fragments do not cluster in one specific and predictable place on the $ZA$-plane---such as a 1/8 fraction of ($Z_0, A_0$), for example---but instead disperse rather broadly and randomly along both $Z$ and $A$ axes.
Here, $A = Z + N$ is the total number of nucleons, and $Z$ and $N$ are the numbers of protons and neutrons, respectively. 

To contrast the fission-driven nucleogenesis with nucleosynthesis: nucleosynthesis creates nuclei by moving {\em upward} along the valley-of-stability (colored cells representing the longest half-lives), 
while fission creates nuclei by jumping all over the white-cells converging onto the valley-of-stability from both sides, moving {\em downward} from extremely high $A$ numbers. 
In this context, a useful reminder is offered in Ref. \cite{Giuliani_2019}: 
«According the report of the Transfermium Working Group \cite{Wapstra_1991}, 
in order to talk about a new element, the corresponding nuclide with
an atomic number $Z$ must exist for at least $10^{-14}$s, which is a reasonable estimate of the
time it takes a nucleus to acquire its outer electrons, bearers of the chemical properties.
Consequently, if for all isotopes of some superheavy element, including isomeric states 
\cite{Heenen_2015,Jachimowicz_2017}, 
nuclear lifetimes are shorter than $10^{-14}$ s, the corresponding element does not exist. 
On the other hand, in order to define a nuclide, its
lifetime should be longer than the single-particle time scale $T_{s.p.} \approx 1.3 \times  10^{-22}$s 
\cite{Goldanskh_1966,Thoennessen_2004} 
that corresponds to the time scale needed to create the nuclear mean field. 
Consequently, there is no chemistry for nuclides with lifetimes between $10^{-14}$s and $10^{-22}$s.»
Although there is no chemistry for super-short-lived nuclides, they nonetheless play critically important roles in fission-driven nucleogenetic cascades.

\begin{figure}[H]  
 \includegraphics[width=0.93\columnwidth]{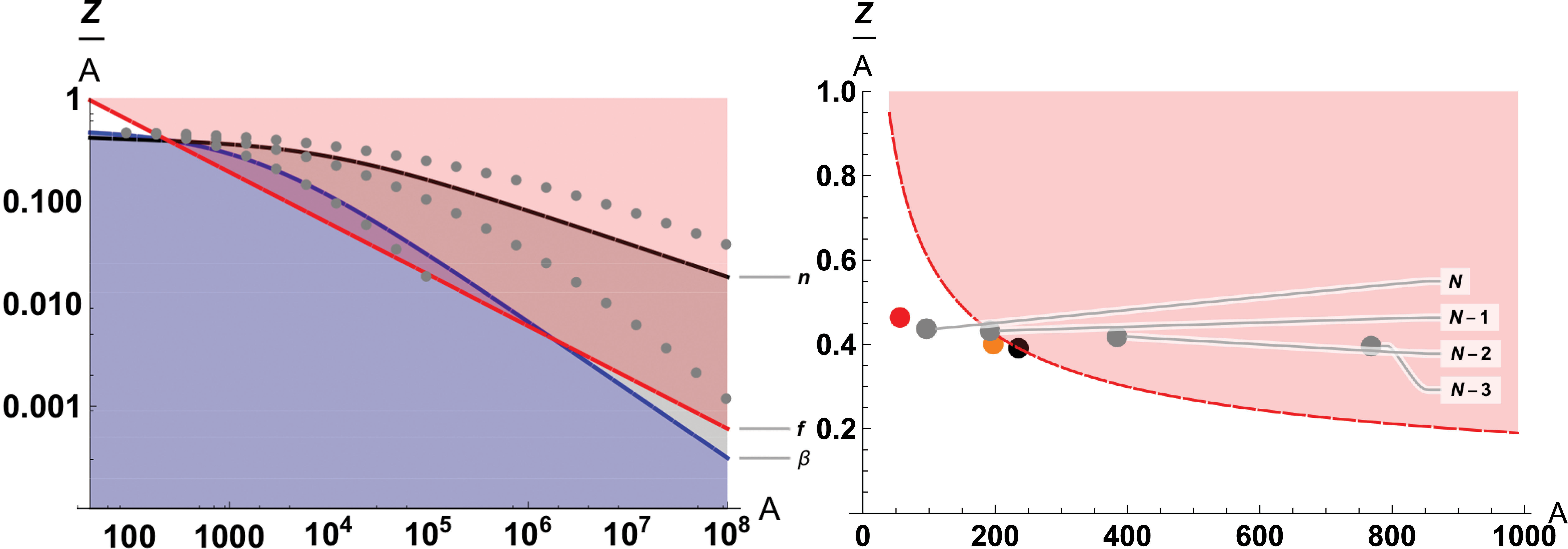} 
\caption
[Fission-Events Framework: Nucleogenetic Cascades (Fission Paths from Mega-Nuclei)]
{
{\bf Both Panels:} 
Evolution ``jumps'' in  fission cascades (from right to left), 
starting from an initial giant nucleus 
($A \sim$$10^8$) and, in a couple of dozens of splits (gray dots),  
 leading to $^{92}{\rm Mo}$ (final left gray dot). 
For each ``daughter''-nucleus, the ``ancestor'' simply splits in two (and a parametrized adjustment for the ``loss'' of neutrons due to $n$- and $\beta$-emissions is included at each fission event). 
Critical lines (where binding energy $B$ have extrema, i.e., $(\partial_iB)_j=0$ with indices $i,j$ denoting nucleon numbers $Z, N, A=Z+N$) are depicted as: $\beta$-line (blue), $n$-line (black), and fission-limit $f$-line (red). 
(see Ref.~\cite{Tito_2020} for derivations and details.)  
In the shaded red zone, fission takes place.
In the shaded gray zone, $n$-emission takes place.
In the shaded purple zone, $\beta$-emission takes place. 
{\bf Right Panel:}
Zoom into $A<1000$ only (not logarithmic but liner scale): the {\em final} stage of the fission-cascade leading to $^{92}{\rm Mo}$. 
Gray dots show  $^{92}{\rm Mo}$ (label N) and its last three generations of ``ancestors'' (N-1, N-2, N-3). 
Placements of {\rm Fe} (red dot), {\rm Au} (yellow dot), and {\rm U} (black dot) are \mbox{also depicted}.
}
\label{Fig:2}
\end{figure}

Figure~\ref{Fig:2} illustrates, using simplified assumptions, how fission can rapidly lead to formation of familiar nuclei (such as $^{92}{\rm Mo}$) from $A \sim$$10^8$ (see Ref.~\cite{Tito_2020} for derivations). 

{\bf Averaging,  Scaling, and Use of Solar Pattern as Benchmark:}    
In analyses of stellar data, 
(for example, when studying actinide-rich and actinide-poor $r$-process-enhanced metal-poor stars), 
procedures similar to the one described in Ref. \cite{Holmbeck_2019} 
are typically used (emphasis added): 
\begin{quote}
 «We treat these three levels of relative actinide enhancement
as three distinct “groups” and assume that each group’s
members formed from gas enriched by an individual $r$-process
event.
\dots 
Within each group, we consider the relative variations among the limited $r$-elements as well as the actinides as intrinsic to the progenitor $r$-process event. \dots

«We {\em combine the abundances} of stars within 
[each group]
by scaling the individual abundance patterns to the respective average residual obtained from comparison with the Solar $r$-process pattern between $_{56}{\rm Ba}$ and $_{71}{\rm Lu}$.  After scaling the solar pattern such that the {\em average
deviation of the stellar pattern from solar pattern between {\rm Ba} to {\rm Lu} is minimized}, we find the range of scaled abundances derived for each element \dots »  
\end{quote}
 
 In the traditional framework, such reasoning makes perfect sense.  
 In the fission-event framework, however, the following considerations intervene: 
First, individual stars may, in theory, be impacted by the fission-events. 
Second, 
according to Ref. \cite{Tito_2020}, 
the Sun itself is an ``impacted'' star. 
Depending on the posed questions, its composition may or may not be the appropriate benchmark.  
Recall also that, for the Sun, what is known as  the ``solar system abundances profile'' is actually 
an  {\em assemblage} of the results of analyses of solar photosphere, solar wind, and the terrestrial and meteoritic data  
 \cite{Lodders_2009,Asplund_2009,Grevesse_2014,Scott_2014a,Scott_2014b,Lodders_2020}. 
According to Ref. \cite{Tito_2020}, the ``rocky'' samples represent the material produced by the fission-event within the inner part of the solar system. 

{\bf Fission-Event Impact on Stellar Spectrum:} 
The fission-event produces enrichment--nuclei 
(on top of the enrichment from all prior galactic sources) 
which may remain in the stellar atmosphere (contributing to spectral signatures) or 
sink into the interior (influencing macroscopic processes).  
This is also relevant because, as well known, many factors affect conversion of raw spectral data into deduced stellar abundances \cite{Hinkel_2016}.   
In particular, the treatment of local thermodynamic equilibrium within stellar atmospheric models has been found to yield dramatically different results for a number of elements \cite{Gehren_2006}. 
NLTE (Non-Local Thermodynamic Equilibrium) seems to be especially important for stars with a low metallicity, or ${\rm [Fe/H]} < -1.0$, where the stellar models deviate from solar \cite{Zhao_2016}. 

Indeed, for the solar system, the Standard Solar Models (SSM) predictions for the sound speed near the core, the surface helium abundance, and neutrino fluxes have \textls[-35]{remained severely discrepant with helioseismological measurements 
(see, for \mbox{example, \cite{Asplund_2009,Serenelli_2009,Villante_2010,Vagnozzi_2017,Vagnozzi_2019})}}.  
In the traditional framework, it has been concluded that: 
«The reason is to be searched for within the huge increase in the abundance of refractory elements 
({\rm Mg, Si, S, Fe}), which leads to a hotter core» \cite{Vagnozzi_2019}. 
In the fission-events framework, as discussed in Ref. \cite{Tito_2020}, this so-called ``solar modeling problem'' may simply indicate that the Sun's interior in fact contains some enrichment-material from the fission-event. 

{\bf Observable ``Impact Signatures'':} 
An important test-question may be asked:  If looked at from afar, could the solar spectrum reveal that the Sun as an ``impacted'' system,  among presumably ``pristine'' neighbors? 
If so, can the insights be used for spotting other ``impacted'' systems? 

One indication has been noted for awhile: abundances of a number of 
elements---such as {\rm Fe, C, N}---tend to be somewhat higher in the Sun than in the B stars 
\cite{Asplund_2009}.   
In the traditional framework, various explanations have been contemplated, but it was concluded that: 
«It is unclear whether the solution can be found in the solar or B star analyses or, if a real difference indeed exists, perhaps due to infall of low-metallicity gas to the solar neighborhood»  
 \cite{Asplund_2009}.
In the fission-event framework, this ``excessive'' enrichment of the Sun can be explained as the result of the ``in-system'' fission-event. 
Perhaps other systems with ``excessive''  {\rm Fe, C, N}  may be also considered candidates for being  ``impacted''.  

Another possible suggestion is to examine more thoroughly the elements with $p$-isotopes. 
Indeed, as discussed in Ref. \cite{Tito_2020}, one strong evidence (among many) in support of the conclusion that the solar system is in fact an ``impacted'' system, is the presence of $p$-isotopes in terrestrial and meteoritic samples. 
Unlike $r$- and $s$-nuclei, $p$-nuclei {\em cannot} be produced by supernovae---they must have come from some other source. 
In the two-stage 
fission-event framework, 
these isotopes were produced in the solar system by the ``in-system'' fission-event.  
Overall, over thirty $p$-isotopes have been identified in the meteorites of the solar system: 
 ${\rm ^{74}Se}$, 
  ${\rm ^{78}Kr}$, 
   ${\rm ^{84}Sr}$, 
    ${\rm ^{92}Mo}$, 
     ${\rm ^{94}Mo}$, 
      ${\rm ^{96}Ru}$, 
       ${\rm ^{98}Ru}$, 
        ${\rm ^{102}Cd}$, 
         ${\rm ^{106}Cd}$, 
          ${\rm ^{108}Cd}$, 
           ${\rm ^{113}In}$, 
            ${\rm ^{112}Sn}$, 
             ${\rm ^{114}Sn}$, 
              ${\rm ^{115}Sn}$, 
               ${\rm ^{120}Te}$, 
                ${\rm ^{124}Xe}$, 
                 ${\rm ^{126}Xe}$, 
                  ${\rm ^{130}Ba}$, 
                   ${\rm ^{132}Ba}$, 
                    ${\rm ^{138}La}$, 
                     ${\rm ^{136}Ce}$, 
                      ${\rm ^{138}Ce}$, 
                       ${\rm ^{144}Sm}$, 
                        ${\rm ^{152}Gd}$, 
                         ${\rm ^{156}Dy}$, 
                          ${\rm ^{158}Dy}$, 
                           ${\rm ^{162}Er}$, 
                           ${\rm ^{164}Er}$, 
                           ${\rm ^{168}Yb}$, 
                           ${\rm ^{174}Hf}$, 
                           ${\rm ^{180}Ta}$, 
                           ${\rm ^{180}W}$, 
                           ${\rm ^{184}Os}$, 
                           ${\rm ^{190}Pt}$, 
                           ${\rm ^{196}Hg}$ 
                           \cite{Rauscher_2013}.     
Except for ${\rm ^{92}Mo}$ and ${\rm ^{94}Mo}$ (14.77\% and 9.23\% of total {\rm Mo}) and ${\rm ^{94}Ru}$ (5.54\% of total {\rm Ru}), their relative abundance is less than 2\% of the respective element. 
In comparison with the more neutron-rich isotopes, the $p$-nuclei are typically 10--1000 times less abundant. 
At present, massive stars are thought to produce $p$-nuclei through photodisintegration of pre-existing intermediate and heavy nuclei. This so-called $\gamma$-process requires high stellar plasma temperatures and occurs mainly in explosive {\rm O/Ne} burning during a core-collapse supernova. 
Although models of  the $\gamma$-process in massive stars have been successful in producing a large range of $p$-nuclei, significant deficiencies remain \cite{Rauscher_2013}.  
As noted in Ref. \cite{Pignatari_2016}: 
«After more than fifty years of research, the production of the $p$ nuclei still carries several mysteries and open questions that need to be answered.»
Furthermore, in the meteoritic samples, $p$-nuclei have been found mixed-in with other nuclei in such ways which make it difficult to reconcile with the traditional distant-supernovae explanations.  
For example, some so-called presolar nano-diamonds carry the so-called  {\rm Xe}-HL \mbox{component \cite{Lewis_1987}}.  
The {\rm Xe}-HL signature is made by enhanced light and heavy stable nuclei: ${\rm ^{124,126}Xe}$ ({\rm Xe}-L) and ${\rm ^{134,136}Xe}$ ({\rm Xe}-H). 
In the traditional framework, ${\rm ^{124,126}Xe}$ are the $p$-nuclei believed to be produced by the $\gamma$-process, while ${\rm ^{134,136}Xe}$ are believed to be formed by the $r$-process in core collapse supernovae, condensing in CCSNe ejecta. 
However, in the diamond samples, {\rm Xe}-L cannot be disentangled from {\rm Xe}-H since the diamonds carrying the two components are well mixed. 
As stated by Ref. \cite{Pignatari_2016}: 
«The corresponding process cannot be explained so far. 
Furthermore, diamonds are carbon-rich grains, while the $\gamma$  process is activated in oxygen-rich stellar layers, where carbon-rich dust should not form [...] 
Last but most importantly, the isotopic ratio of the {\rm Xe}-L isotopes is not consistent with the same ratio of these $p$ nuclei in the solar system [...] 
It is unknown why they are different.» 
In the fission-event framework, the problem is resolved because all nuclei are produced ``locally'' in the inner-part of the solar system as a result of the fission-event. 
Thus, generally speaking, the presence of the $p$-nuclei detected in the solar system's ``rocky'' material appears to be the evidence indicating that the solar system is likely an ``impacted'' stellar system. 

For other stellar systems, unfortunately, there are no meteoritic samples to study, and   
not even all elements or isotopes can be distinguished in spectra. 
However, among the elements with $p$-isotopes, {\rm Mo} is ``special'' because terrestrial and meteoritic samples reveal that a substantial  portion of {\rm Mo} in the solar system ($\sim$$24\%$) is in its $p$-isotopes. 
Although stellar spectral analyses do not distinguish  {\rm Mo}-isotopes, if it could be presumed that ``impacted'' systems may also contain a substantial share of  $p$-isotopes among all {\rm Mo}-isotopes, 
then the spectra of impacted systems may perhaps show ``excess'' of total {\rm Mo} relative to elements with no $p$-isotopes---for example, relative to {\rm Fe}.

\begin{figure}[H]  
 \includegraphics[width=0.50\columnwidth]{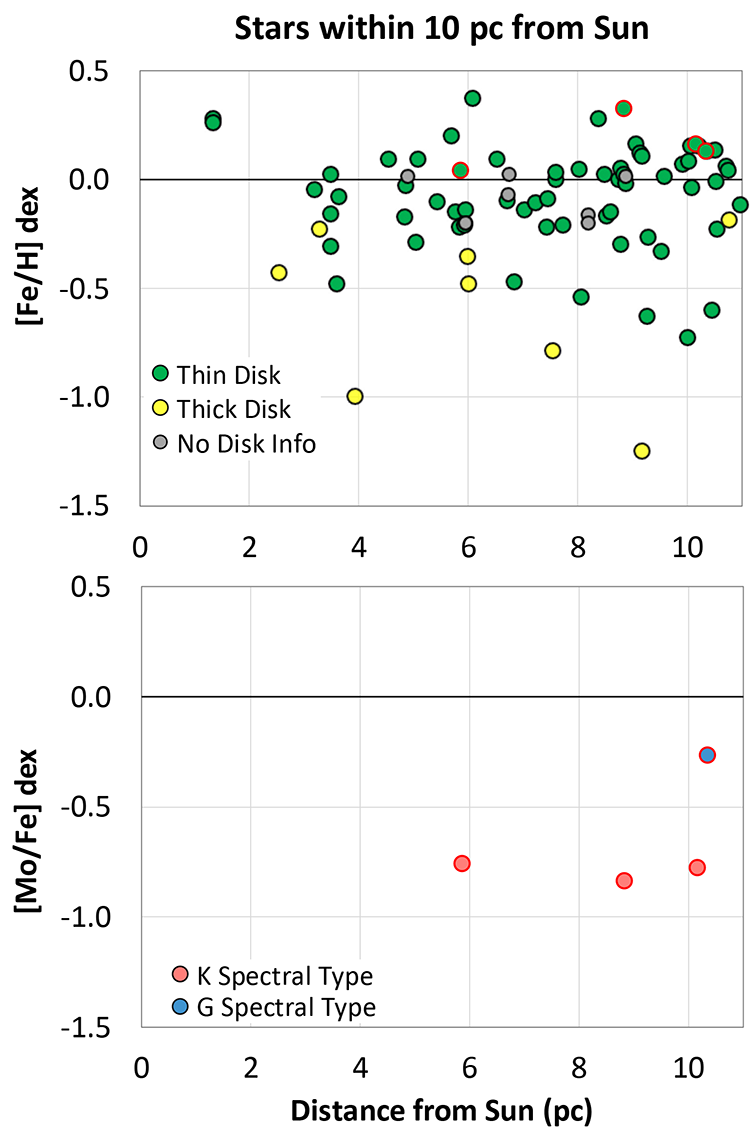}
\caption
[Stellar Systems: {\rm [Fe/H]} and  {\rm [Mo/Fe]} Ratios for Stars within 10~Parsec from Sun.] 
{
Ratios {\rm [Fe/H]} ($\equiv$ metallicity; upper panel) and  {\rm [Mo/Fe]} (lower panel) 
for the stars within $\sim$10~parsec from the Sun. 
R ed rim marks those stars which appear in both panels. 
Data from the Hypatia Catalog \cite{Hinkel_2014}, accessed on 30 November 2020. 
Notation [A/B] describes the relative abundances of two elements in a star compared to that in the Sun: [A/B] = log$_{10}(N_A/N_B) - $log$_{10}(N_A/N_B)_\odot$. 
A star with {\rm [Fe/H]} = $-1$, for example, contains a factor $10^1$ fewer {\rm Fe} 
nuclei by number than the Sun. 
The associated unit is a logarithmic unit: dex (contraction of ``decimal exponent''). 
} 
\label{Fig:3}
\end{figure}

Illustrating this logic and hypothetically studying the Sun from afar, 
the presumably ``pristine'' neighbors should seem ``depleted'' in {\rm Mo} relative to the ``impacted'' Sun. 
\mbox{Figure~\ref{Fig:3}} shows {\rm [Fe/H]} ratios (upper panel) and {\rm [Mo/Fe]} ratios (lower panel)  for the stars in the $\sim$10~parsec solar neighborhood.  
Since stars can migrate or scatter into or out of the solar neighborhood, 
$10$~parsecs may perhaps be considered the ``close'' neighborhood which presumably changed to a lesser degree than the ``greater'' neighborhood. 

While the {\rm Fe} data are available for numerous stars within $\sim$10~parsec from the Sun (Figure~\ref{Fig:3}, upper panel),   
the {\rm Mo} data are available only for four stars.  
The three closest stars 
indeed show {\rm [Mo/Fe]}~$\sim$$-0.8$ (Figure~\ref{Fig:3}, lower panel).  
In other words, if---as traditionally presumed---all the stars in the neighborhood, including the Sun, formed from the nebulas that were similarly enriched by the {\em same distant cataclysms}, the neighbors would be similar to the Sun and would show up on the plot with {\rm [Mo/Fe]}~$\simeq 0$. 
However, the Sun appears to be more {\rm Mo}-enriched than its closest neighbors, 
while the three neighbors are similar to each other. 
This is consistent with the premise that a candidate for being an impacted system may perhaps be noticed among its pristine neighbors by the ``excessive'' {\rm Mo} abundance.

{\bf Chronometers:}
The possibility of ``in-system'' fission-events carries great implications for chronometric estimations. 
Long-lived isotopes ${\rm ^{232}Th}$ and ${\rm ^{238}U}$---with half-lives of 14~Gyr and 4.5~Gyr, respectively---are used for age measurements  of stellar objects because these half-lives cover cosmic timescales.  
(However, while absorption lines of {\rm Th} are regularly measured,  a {\rm U} detection is very difficult because only one, extremely weak, line is available in the optical spectrum.)
As described in Ref. \cite{Jacobson_2014}, 
three types of element combinations involving radioactive and naturally occurring stable elements,  abbreviated with “s” in the following equations, are used as  chronometers: 
\begin{quote}
$\Delta t = 46.7 {\rm [log (Th/s)_{initial} - log} ~ \epsilon {\rm (Th/s)_{now} ] } $, 

$\Delta t = 14.8 {\rm [log (U/s)_{initial}  - log} ~ \epsilon {\rm  (U/s)_{now} ] } $, 

$\Delta t = 21.8 {\rm [log (U/Th)_{initial}  - log} ~ \epsilon {\rm  (U/Th)_{now} ] } $. 
\end{quote}
\vspace{12pt}

Here, the subscript “initial” refers to the theoretically derived initial production ratio of these elements, while the subscript “now” refers to the observed value of the abundance 
($\epsilon$)  
ratio. 
Because, traditionally, it has been assumed that the $r$-process is responsible for the production of  {\rm Th} and  {\rm U}, stable  $r$-process elements (such as {\rm Eu, Os,} and {\rm Ir}) 
are typically used in the estimations.  
However, since the  {\rm U/Th} chronometer was first measured in CS 31082-001 \cite{Hill_2002}, 
it has been noted that, for the stars which suffer from the “actinide \mbox{boost” \cite{Honda_2004}}, 
the {\rm Th/Eu} ratios yield negative ages.   
(The “actinide boost”  means that, compared with the scaled solar $r$-process, they contain too much  {\rm Th} and  {\rm U}). 
According to Ref. \cite{Jacobson_2014}: «The origin of this issue has yet to be understood». 
As noted earlier, about a quarter of strongly $r$-process enhanced stars shows {\rm Th} abundances that are higher than expected compared to other stable $r$-process elemental abundances and the scaled solar $r$-process pattern. 
«One explanation may be that these stars show the $r$-process pattern of two $r$-process events that occurred at different times---one just prior to the star’s formation and one at a later time in the vicinity of the star»   \cite{Jacobson_2014}.  

The fission-event framework offers a solution.  
Within the framework,  
the heavy elements are indeed produced by the processes occurring «at different times---one just prior to the star’s formation and one at a later time in the vicinity of the star» \cite{Jacobson_2014},   
with the fission-event being the latter. 
Furthermore, the possibility of fission-cascades occurring {\em ``in-system''}  
is fundamentally important for chronometer-calculations  because of their two built-in assumptions:  
(1) the system is presumed to be in isotopic equilibrium ({\em homogeneous, uniform}) at time $t = 0$, 
and 
(2) the system as a whole and each analyzed part of it is presumed to be 
 {\em closed} during the entire time. 
Violations of these conditions are the principal sources of errors in chronometer-calculations. 

\section{Exoplanetary Systems}
\label{s:4}

As the expansive and growing database of exoplanetary systems is revealed:  
«Extra-solar planetary systems do not typically look like the Solar System» \cite{Raymond_2014}. 

\begin{figure}[H]  
\includegraphics[width=0.80\columnwidth]{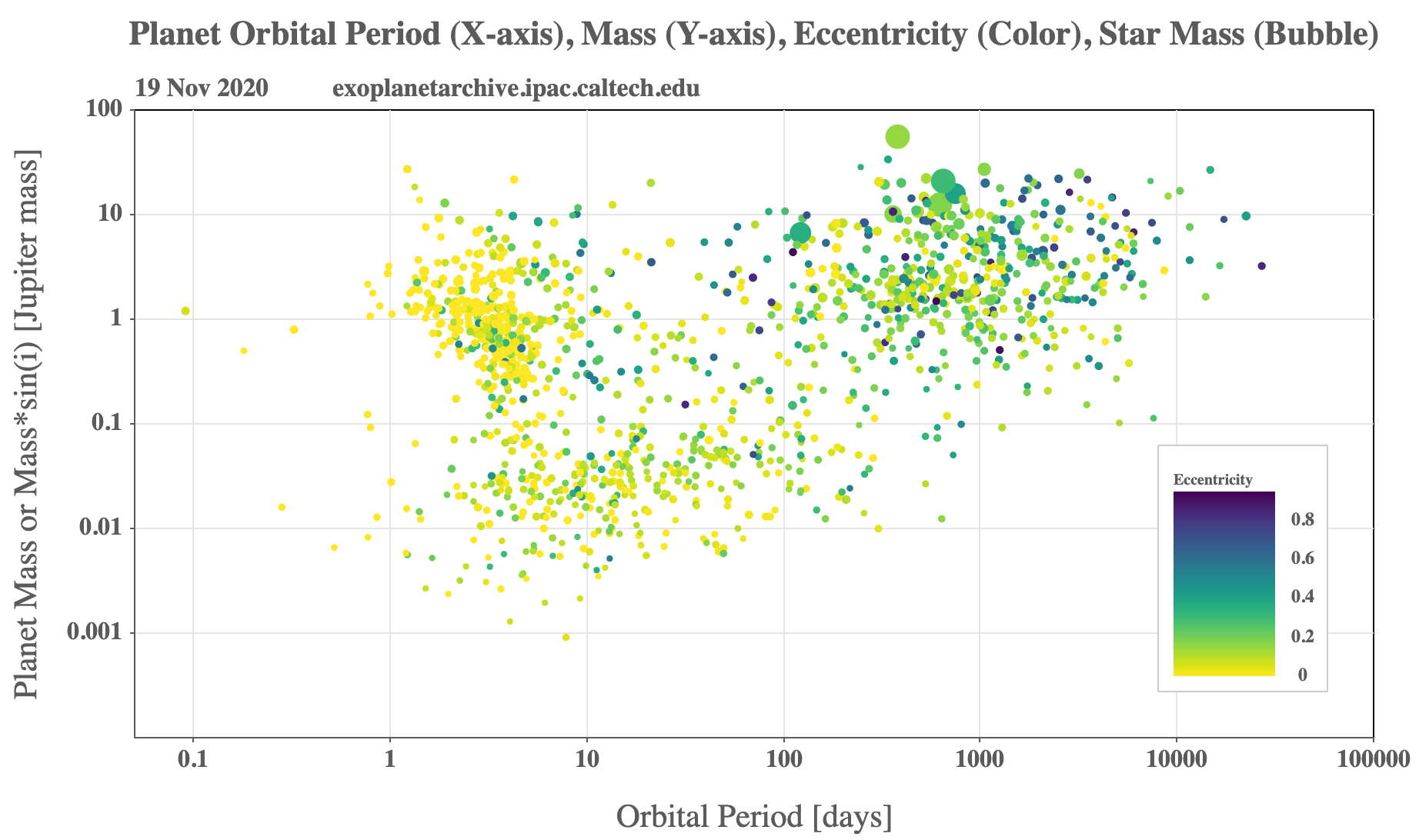}
\caption
[Exoplanets: Orbital Period, Planet Mass, Host Star Mass, Eccentricity] 
{
 Orbital characteristics of exoplanets in \citet{Exo_Confirmed}.      
}
\label{Fig:4}
\end{figure}

One notable peculiarity of many exoplanets, as can be seen in Figure~\ref{Fig:4}, is that they have very short orbital periods---just several days---implying that they orbit extremely closely to their host stars, much closer than the solar system's innermost planet Mercury whose period is 88 days. 
Some of such planets have low eccentricities, some have high. 
The presence of giant planets close to their host stars is challenging to explain. 
At present, they are thought to have either undergone extensive inward gas-driven migration or been re-circularized by star-planet tidal interactions from very eccentric orbits produced by planet-planet scattering or other mechanisms \cite{Raymond_2014}.

Another peculiarity of many exoplanets is their high density. 

\begin{figure}[H]  
\includegraphics[width=0.95\columnwidth]{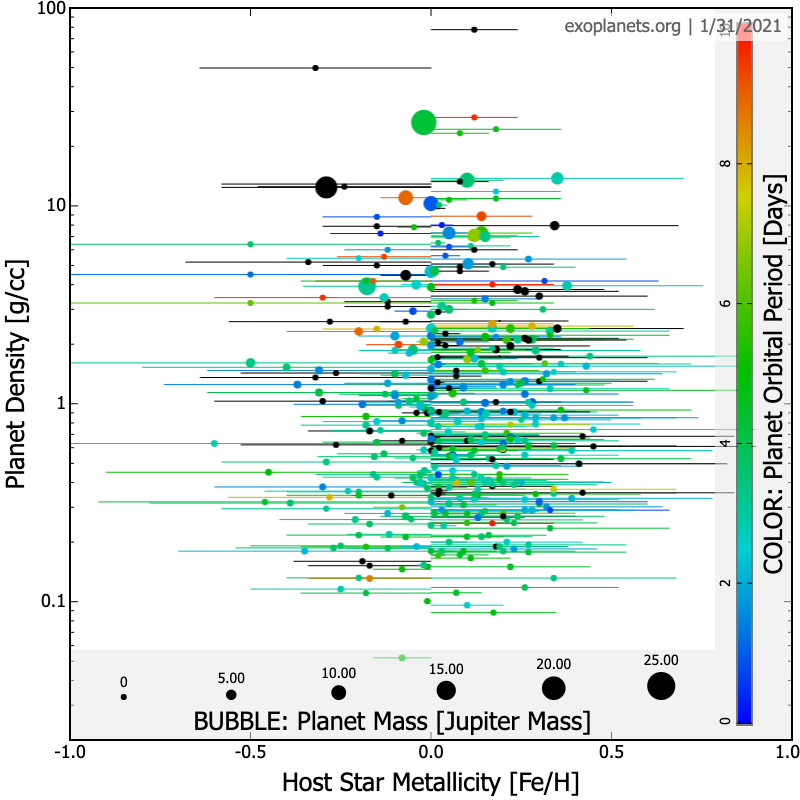}
\caption
{
 Exoplanet densities ($y$-axis), masses (bubble-sizes), orbital periods (color), 
and their host--star metallicities ($x$-axis). 
Star metallicity error-bars are depicted, but planet density error-bars are omitted for visual clarity. 
Black colors represent planets with orbital periods greater than 10 days. 
Color scale is chosen to maximize visual clarity.  
Data from the Exoplanet Orbit Database \cite{Han_2014}.  
}
\label{Fig:5}
\end{figure}

Before discussing this issue in greater detail, it may be noted first that, intuitively, one would expect to see some correlation between densities of the planets and metallicities of their host stars. 
However, the data show no such correlation yet.  
Figure~\ref{Fig:5} 
plots planet-density versus host-metallicity [Fe/H].   
Error-bars represent the uncertainties in the host--metallicity estimates. 
Error-bars for density estimates are not plotted for visual clarity.  
Bubble sizes add information about planet masses and 
colors---about planet orbital periods (black denotes planets with orbital periods greater than 10 days). 
Hopefully, further advancements in the spectral observations and data analyses may 
bring better insights into whether any correlation indeed exists or not. 

\begin{figure}[H]  
     \includegraphics[width=0.48\columnwidth]{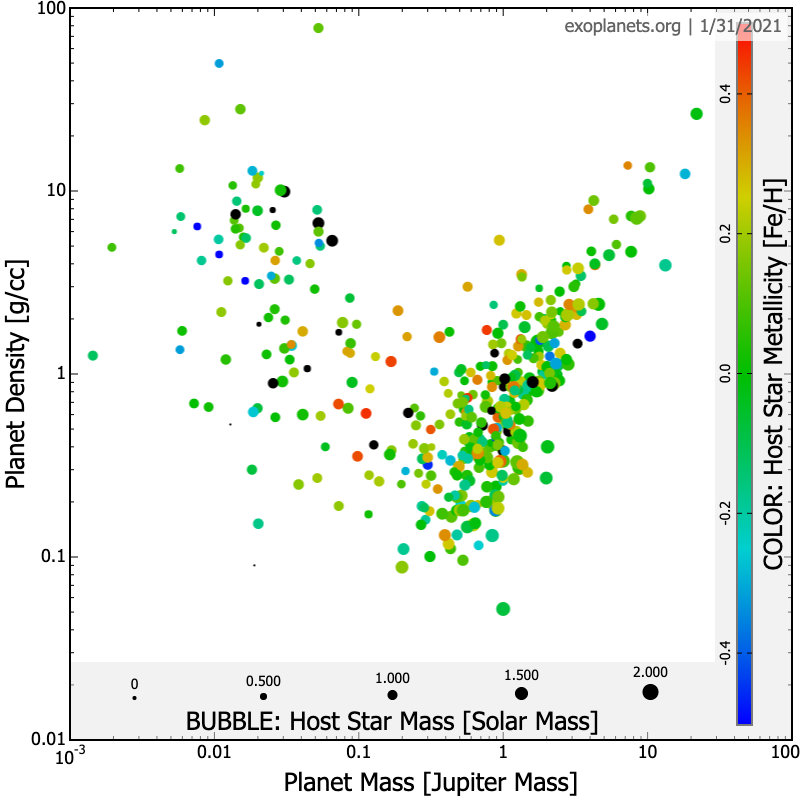}
 \hspace{0.01\textwidth}
  \includegraphics[width=0.48\columnwidth]{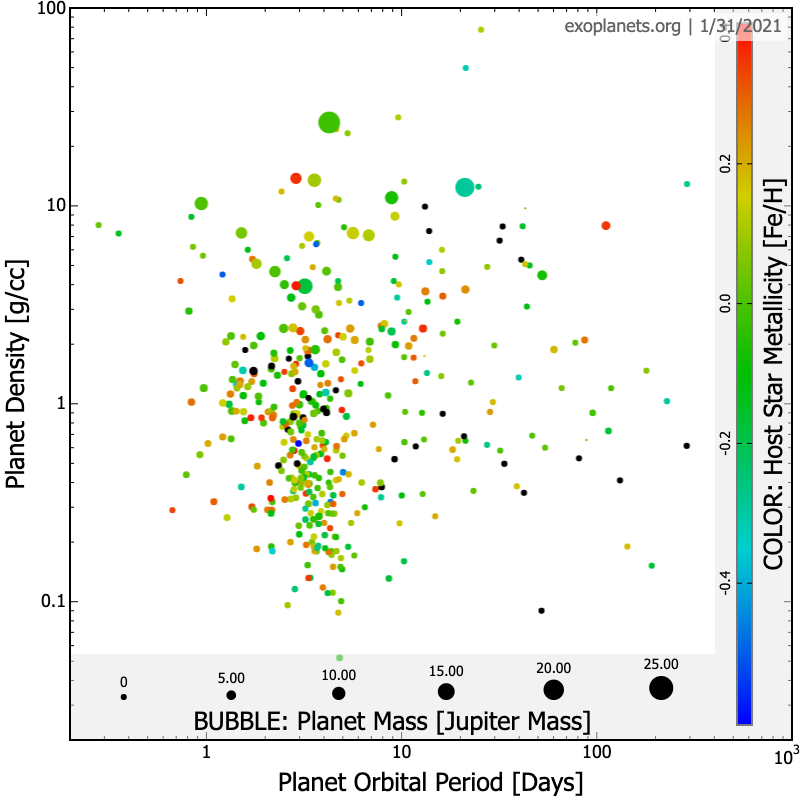}
      \caption{
 Exoplanet characteristics: planet density (\textbf{Both panels}, $y$-axis), planet mass (\textbf{Both Panels}, $x$-axis or bubble-size), planet orbital period (\textbf{Right Panel}, $x$-axis), host--star metallicity [Fe/H] (\textbf{Both Panels}, color), and host--star mass (\textbf{Left Panel}, bubble-size). 
Source: the Exoplanet Orbit \mbox{Database \cite{Han_2014}}. 
}
\label{Fig:6}
\end{figure}

Furthermore, Figure~\ref{Fig:6} (left panel) shows that, for massive planets (Jupiter-size and greater), there seems to exist some correlation between planet density and planet mass, but for less massive planets no correlation is apparent. 
By plotting orbital periods on $x$-axis, Figure~\ref{Fig:6} (right panel) reveals the planets' 
distances from their host stars more clearly, but shows no apparent correlations.   

\clearpage


Recall that the bulk densities of the solar system planets 
(cited in g/cc, grams per cubic centimeter) 
are: 
Mercury $ \sim$5.4, 
Venus $ \sim$5.2, 
Earth $ \sim$5.5, 
Mars $ \sim$3.9, 
Jupiter $ \sim$1.3, 
Saturn $ \sim$0.7, 
Uranus $ \sim$1.3, 
Neptune $ \sim$1.6. 
The bulk density of the Sun is 1.4~g/cc.
Thus, as Figures~\ref{Fig:5} and~\ref{Fig:6} have shown, there exist 
a number of exoplanets with bulk densities significantly exceeding the densities of all planets of the solar system;  
such planets have to be presumed to be significantly enriched with heavy elements. 

In this context, 
another useful reminder is about which elements are considered ``heavy'' or ``metal'' in different fields. 
In stellar observations, ``metallicity'' refers to the proportion of the material that is in elements other than {\rm H} and {\rm He}, so even {\rm Li} is a ``metal''. 
(However, as mentioned earlier, {\rm Fe} is typically used as a proxy for metallicity because the large number of {\rm Fe} absorption lines present in the optical wavelength regime makes it straightforward to measure \cite{Jacobson_2014}.)
In nuclear physics, however, the ``heavy'' elements are the post-{\rm Fe}-group elements---they are produced via nucleosynthesis in supernovae, in neutron-star mergers, and during fission of super-heavy and super-super-heavy elements (as discussed in \cite{Tito_2020}). 
Post-{\rm Fe} elements have much greater densities than $\sim$5.5~g/cc (the bulk density of Earth).  
For example: 
{\rm Fe} $\sim$ 7.9, 
{\rm Co $\sim$ 8.9, 
{\rm Bi} $\sim$ 9.8, 
{\rm Pb} $\sim$ 11.4, 
{\rm Hg} $\sim$ 13.5, 
{\rm U} $\sim$ 18.9, 
{\rm Au} $\sim$ 19.3, 
{\rm Pt} $\sim$ 21.5, 
{\rm Os} $\sim$ 22.6~g/cc. 

Figure~\ref{Fig:7} summarizes the details of density estimates for exoplanets with bulk density $\rho > 5$~g/cc. 
%
When multiple density estimates are available for a planet, in Figure~\ref{Fig:7}, they 
are clustered together; one label---the planet's name---is placed above each cluster.  
The clusters are ordered so that planet masses increase along the $x$-axis. 
The planets on the left side of the chart have masses ranging from $\sim$0.01~$M_{Jup}$ to $\sim$0.1~$M_{Jup}$, while the planets on the right side---from $\sim$5 to $27~M_{Jup}$. 
The magnitudes of uncertainties (error-bars) reflect the challenges of the estimation process. 
In some studies, it was simply concluded that the planet bulk density was less than some number 
(such planets are excluded from Figure~\ref{Fig:7}), 
but, in many cases, more specific determinations were made. 
For example, the latest (2017) results for planet CoRoT-3b (with $\rho =25.9^{+6.4}_{-4.9}$ g/cc) are explained in Ref.~\cite{Bonomo_2017}. 
For visual comparison, the colored horizontal lines indicate densities of 
iron ${\rm ^{56}Fe}$ ($7.9$~g/cc, red line), 
lead ${\rm ^{82}Pb}$   ($11.4$~g/cc,  blue line),  and 
gold ${\rm ^{79}Au}$  ($19.3$~g/cc, yellow line).  

\begin{figure}[H]  
\includegraphics[width=0.99\columnwidth]{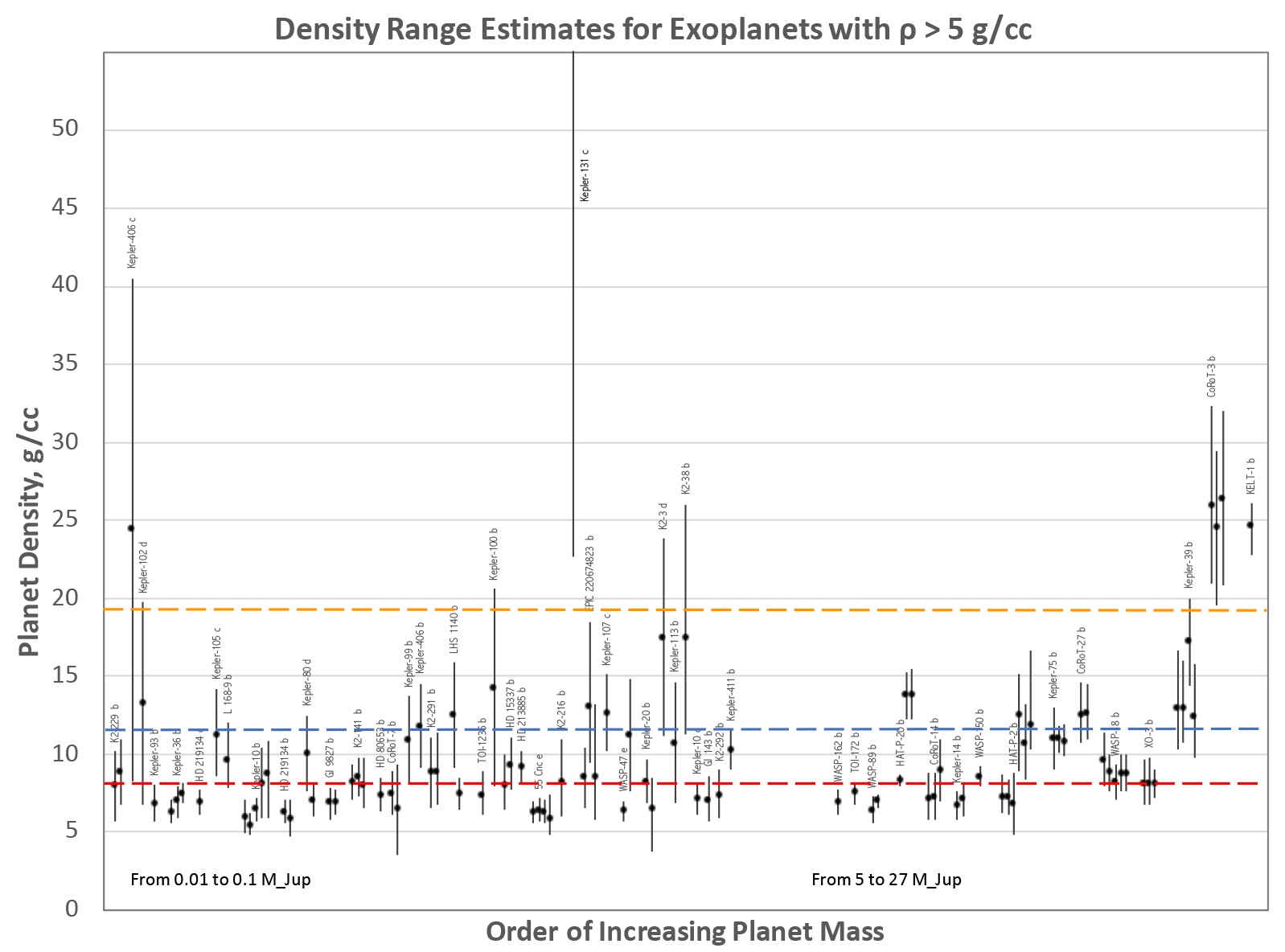}
\caption
[Density of Exoplanets: Details of Range Estimates for Planets with $\rho>5$ g/cc] 
{
 Estimate details for planets with $\rho > 5$~g/cc.  
Data from \cite{NASAESI2020}, accessed on November 28, 2020. 
Colored lines show densities (g/cc): 
red---{\rm Fe} ($7.9$), 
blue---{\rm Pb}   ($11.4$),  
yellow---{\rm Au}  ($19.3$).   
}
\label{Fig:7}
\end{figure}


As  Figure~\ref{Fig:7} shows, a number of planets appear to be composed of highly-dense ``heavy'' post-{\rm Fe} elements.  
In the fission-event framework, their origins are transparent: they were formed from the debris produced by a fission-event. 
In fact, such high exo-densities may perhaps be the pointers to the possible presence of the not-yet-decayed super-heavy elements---perhaps the ones from the theorized ``islands of stability'' 
(see Ref. \cite{Tito_2020}, and references therein, such as~\cite{Greiner_2013,Afanasjev_2018}). 
At present, several super-heavy  ``islands of stability'' have been theoretically predicted: 
(Z$\sim$114,~N$\sim$184--196), (Z$\sim$138,~N$\sim$230), (Z$\sim$156,~N$\sim$310), and~(Z$\sim$174,~N$\sim$410), 
here $Z$ is a number of protons and $N$ is a number of neutrons in a nucleus.
In contrast, within the conventional framework of nuclei-formation via nucleosynthesis, 
the existence of exoplanets with post-{\rm Fe} bulk densities is difficult to explain, creating an impression that high density estimates are erroneous  (when in fact they may be the evidence confirming 
our fission-events hypothesis).  

Indeed, it is generally accepted---see, for example, Ref. \cite{Marov_2020}---that 
the densities of the planets reflect the composition of their protoplanetary disk.  
The disk 
forms from the nebula ($\sim$0.1--1 parsec),
which is a fragment of a giant interstellar molecular cloud spanning tens of parsecs. 
The nebula composition is mostly ($\sim$98$\%$) {\rm H} and {\rm He}, 
but already enriched with ``heavier'' elements formed as the result of nucleosynthesis 
in the stars of previous generations and supernova explosions. 
The sequence of planetary system formation processes includes: 
fragmentation of the interstellar molecular cloud which gradually becomes dense;  
formation of protoplanetary accretionary gas-dust disk surrounding the host--star;  
disk segmentation into initial clumps from which eventually solid bodies (planetesimals), 
planet cores, and planets form. 
The key processes are various types of instability (hydrodynamic and gravitational) in the disk, 
formation of solid bodies, and their subsequent growth. 
Clusters appearing at the initial fragmentation as the result of gravitational instability, initially contain {\em sub-micron-sized} particles (nebula dust and condensates of disk medium). 
 
However, as laboratory experiments have revealed (see Figure~\ref{Fig:8}),  
even small agglomerates do not form easily from dust. 
Most problematic is the range between centimeter- and meter-size bodies.
Even for nano- and micro-meter-sizes (typical for dust of interstellar clouds), dust particle growth 
via van-der-Waals and electrostatic interactions, is \mbox{problematic \cite{Marov_2020}}. 
\vspace{-6pt}
\begin{figure}[H]
\includegraphics[width=0.99\columnwidth]{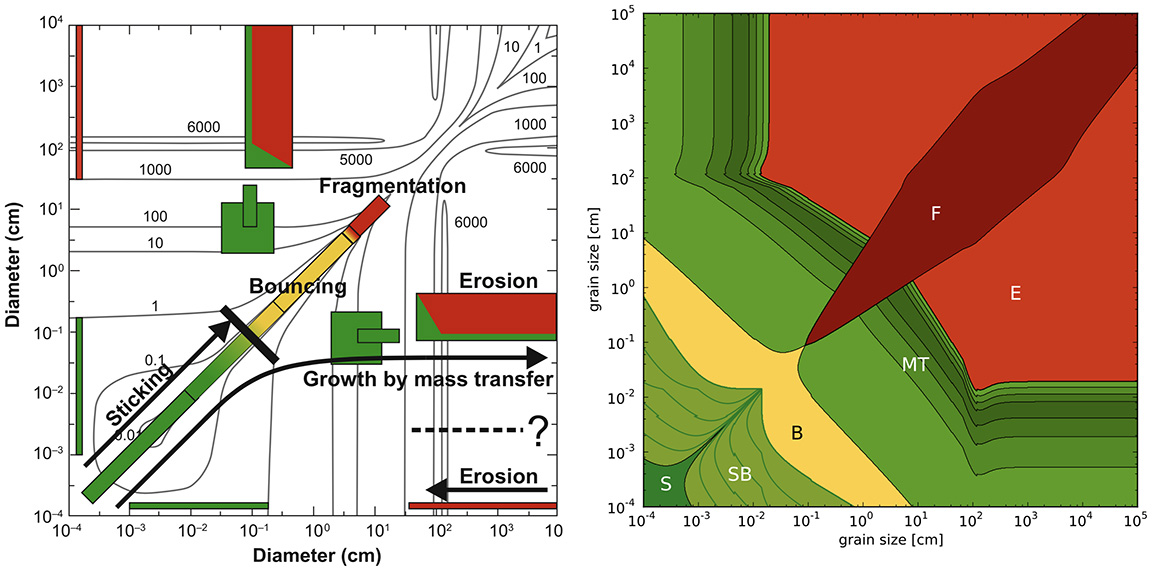}
\caption
[Problem with Dust Accretion in Protodisks: «Sticking Barrier» in Laboratory Experiments]
{
 Schematic representation of the outcomes of dust collisions in protoplanetary disks. 
\mbox{From \cite{Testi_2014}}, reprinted by permission. 
{\bf Left panel}:  the collision velocities (cm/s)  between two dust agglomerates with sizes indicated on the axes in a minimum-mass solar nebula model at 1 AU \cite{weidenschilling1993protostars}. 
The colored boxes denote the explored parameter space and results of laboratory experiments. 
Green represents sticking or mass transfer, yellow bouncing, and red fragmentation or erosion. 
``Sticking'' growth \cite{Zsom_2010}  is prevented by bouncing. 
A possible direct path to the formation of planetesimals is indicated by the arrow ``Growth by mass transfer''. 
{\bf Right panel}: the parameter space for collisions outcomes between bare silicates grains used by numerical models of dust evolution \cite{Windmark_2012}. 
Abbreviations: S--Sticking, B--Bouncing, SB--Sticking and Bouncing, MT--Growth by Mass Transfer, F--Fragmentation, \mbox{E--Erosion}. 
}
\label{Fig:8}
\end{figure}

The combination of these factors makes it rather difficult to explain in the conventional framework 
how the high-density planets formed. 
The nebula had to contain enough post-{\rm Fe} micro-dust for the entire denser-than-iron planet.    
(Recall that, in the solar system, such highly dense elements as {\rm Au} or {\rm Os} are over six orders of magnitude less abundant than {\rm Fe}; see, for example, \cite{Lodders_2020}).
If supernovae were the sources of such enrichment, their ejecta had to be powerfully enriching but somehow non-demolishing the entire nebula in the process, or the highly-dense micro-dust particles had to somehow ``separate'' themselves from the rest and ``aggregate'' (more than they did in the solar system). 
These dust particles also had to overcome the ``sticking'' barrier problem and form the highly-dense planet so large that we can observe it from afar. 

The fission-event framework offers reconciliation of the challenges. 
Indeed, as known for the solar system, 
analyses of more than 400 collected pieces of the Sikhote--Alin meteorite have revealed that the pieces   
 were {\em internally-uniform} chunks composed of ${\rm 88\% \, Fe, 5\% \, Ni}$, and ${\rm 2\% \, Co}$   \cite{Scott_2013,Caporali_2016,Plavcan_2012},  
collectively weighing more than 23~tons, with the largest individual piece being 1745~kg.  
According to our proposal in Ref. \cite{Tito_2020}, the solar system is an ``impacted'' planetary system and some of the known meteorites (especially those classified as irons) may represent the debris from the fission-event. 
Once such large-scale (non-dust-like) debris pieces are formed during the event, larger bodies grow subsequently just as the existing accretion models suggest.  
(In their simulations, the models can simply start with the particle size distributions which include post-meter sizes.)   
Thus, for the planetary systems with highly-dense planets (small or giant), 
fission-events may be the explanation of the origins of such planets. 
In other words, 
a discovery of a high-density planet in a system 
makes the system a candidate for being considered ``impacted''.  

{\bf Planets near Exotic Stars:} 
Finally, it is worth noting that the fission-events framework helps explain the existence of planets around neutron stars. 
Indeed, as well known, the very first exoplanets ever discovered were the planets with masses of at least $2.8~M_{\bigoplus}$ and $3.4~M_{\bigoplus}$ (where $M_{\bigoplus}$ is the mass of the Earth) detected orbiting an old ($\sim$$10^9$~yr) rapidly rotating neutron star (pulsar). 
The planets'  respective distances from the pulsar are 0.47 AU and 0.36 AU, and they move in almost circular orbits with periods of 98.2 and 66.6 \mbox{days 
\cite{Wolszczan_1992}. }
Since then, more planets near pulsars have been discovered. 
At present, formation of planets around post-supernova stars is presumed to depend on destruction of a companion and capture of its material, and therefore is estimated to be rare  \cite{Martin_2016}.
Perhaps focused astronomical observations could offer more insight into such systems and planets. 
Most of the stars presently known to have planets are similar to the Sun   
(main sequence stars of spectral class F, G, or K).
For obvious reasons, such stars attracted most of the initial \mbox{search effort. }
 
\section{Conclusions}
\label{s:5}
 
In this review-paper, we presented an expansion of the ``in-system'' fission-event framework
---which we recently advocated for the solar system  \cite{Tito_2020}---to  exo-systems in general. 
We focused on illustrating the discussion with the already-existing observational data.  

The ``in-system'' fission-events represent a {\em new mechanism} of nucleogenesis in galaxies, 
which may explain certain ``excessive'' or ``peculiar'' abundance patterns in individual stars which 
traditional models struggle to explain at present.      
Therefore, the fission-events framework offers an expansion of the general conception of galactic enrichment. 
The fission-event mechanism and traditional mechanisms differ in two key ways: 

\begin{enumerate}[leftmargin=*,labelsep=4.9mm]

\item	Traditionally, the nuclei are believed to form via {\em nucleosynthesis} (with ``add-on'' decays and fission-(re)cycling), so the general direction of nuclei-creation is {\em upward} (from lower nucleon numbers $A$ to higher $A$). 

In the fission-event framework, the direction is reversed: the nuclei are formed via  fragmentation and {\em fission-cascades} (with ``add-on'' captures and decays), so the general direction of nuclei-creation is
{\em downward}. Fission-cascades start from  giant nuclei with $ln~A~\gg~1$ and include even extremely-short-lived nuclei, and should be distinguished from the {\em upward}-nucleosynthesis-driven fission-(re)cycling used in traditional models which operate in a limited $A$-domain only, disregarding very-short-lived nuclei and nuclei with $A>A_{limit}$ (currently $A_{limit} \sim$320). 

Nucleogenetic signatures of the processes are different, and, therefore, various detected but not-yet-explained  ``excesses'' in stellar element abundances (relative to the traditional models) may have their origins in the fission-events. 
Therefore, spectral observations (and derived isotopic abundances) of multiple ``impacted'' stars may perhaps help nuclear-physics theoreticians by offering factual data for development of constraints for the ``production signature'' of fission-driven nucleogenesis. 
See Ref. \cite{Tito_2020} and references therein, for discussion about the challenges of experimental studies of super-heavy elements in terrestrial conditions.)

\vspace{6pt}  

\item	
Traditionally, it is believed that the ``enrichment material'' is (a) first synthesized {\em at the production-site} and in the ejecta of some cataclysm (such as supernova, AGB star, neutron star merger),  
and (b) then it travels some distance in the form of {\em dust} gradually polluting interstellar gas and encountered nebulas which later form stars and planets. 

In the fission-events framework, the order is reversed: the ``enrichment material'' (a) first travels great distances in the form of a {\em compact drop-like super-dense nuclear-matter object},  
and (b) it explodes in fission-cascades {\em at the encountered stellar system (``in-system'')} 
or in the interstellar space. 

This means that the fission-event ``enrichment material''---being at first a compact drop-like super-dense ``clump'' torn away and catapulted by a supermassive black hole---may travel {\em much greater distances} than the traditionally presumed supernova-ejected ``dust'', thus enriching much more distant regions, not just the local neighborhood of the cataclysm.  Furthermore, the dispersion of fission-event debris (post-explosion) naturally has a relatively ``localized'' character, which may explain some cases of {\em spatial heterogeneity} among stars in clusters or {\em individual outliers} in neighborhoods.  

\end{enumerate}

The fission-events framework implies that all stellar systems may be divided into ``{\em pristine}'' or ``{\em impacted}.'' 
Such classification may be useful when analyzing exoplanetary systems or understanding stellar clusters. 

From the perspective of astronomical observations, 
if an individual stellar system is ``impacted'' by a fission-event, its characteristics 
may be altered in ways which could be detected from afar  
and recognized as ``impact signatures''.  Among them are: 
\begin{enumerate}
\item \textls[-35]{high abundances of post-{\rm Fe} elements (actinides and those called $r$- or $s$- \mbox{process elements});} 
\item high abundances of elements with $p$-isotopes; 
\item high bulk densities of exoplanets; 
\item  short orbital periods of exoplanets. 
\end{enumerate}

From the perspective of evolution, 
if a stellar system is an ``impacted'' system, it means that its evolution had (at least) {\em two distinct stages} and  
at the beginning of each stage it was enriched differently. 
Several types of ``impacted'' systems may be envisioned: 
\begin{enumerate}

\item 
Protonebula Stage: 
If the fission-event occurs during the  protonebula stage of evolution, besides enriching the nebula, the event may serve as the trigger initiating the nebula's gravitational collapse. 
Over the course of subsequent stellar-system evolution, the enrichment is likely to become well-mixed and uniformly distributed, and reveal itself in the spectrum of the eventually-formed host--star and in the composition of the planets. 
\vspace{6pt}  

\item 
Protodisk Stage: 
If the fission-event occurs during the protodisk stage of evolution, disk properties---''dust'' sizes and composition, viscosity, opacity, and so on---may increase abruptly but {\em locally}, thus creating {\em spatial heterogeneity} in the enrichment of the subsequently-formed planets and likely influencing the planets' growth and orbital dynamics. The content and size distribution of solid particles strongly influence the disk thermal regime, viscous properties, turbulence flow patterns, disk medium opacity, snow-line locations, chemical transformations in gaseous medium and, ultimately, its evolution including the processes’ dependence on the radial distance from the protostar and the early subdisk formation. 
Numerical simulations can perhaps examine how evolution paths may change for the 
``{\em t-tau}'' systems---those suddenly impacted at time $t$ during the protodisk's lifetime $\tau_{disk}$ (such as $0.2 \tau$, $0.5 \tau$, and so on, with zero-tau being the nebula-impacted system, and one-tau being the system with fully-formed planets).  
\vspace{6pt}  

\item 
Fully-Formed Stage: 
If the star and the first-stage planets had already formed, then the fission-event may occur  directly {\em within} the star, {\em near} the star, or {\em farther away} from the star. 
The within-star impact, if meaningful, would reveal itself in the stellar spectrum. 
If the fission-event occurs not directly in the star, but its debris is spread within the system, then the debris would orbit the star and accrete via collisions {\em without} any protodisk-effects (since the protodisk had already dissipated by then).
Such event may lead to formation of objects analogous to the solar system's asteroid belt or its terrestrial planets, the proximity of which to the host star would depend on the location of the ``explosion'' and the overall dynamics of the combined system's parts.  
Highly-dense planets may be the representatives of this type of outcomes. 
The debris may also enrich the host--star (and thus reveal itself in the stellar spectrum) and the already-existing planets (perhaps altering their orbits in the process).  This is what happened in the solar system, as we advocated in Ref. \cite{Tito_2020}.

\end{enumerate}

Finally, in view of the vastly more abundant amounts of data available for the solar system (which we discussed in greater detail in Ref. \cite{Tito_2020}) than for any other stellar system, we argue that a rather certain conclusion may be made that the solar system is in fact an ``impacted'' system. It is not ``pristine.''
This means that the question---What is the ``benchmark''?---in many analyses no longer has its usual default answer.  

\vspace{12pt}

\authorcontributions{Conceptualization, E.P.T. and V.I.P.; writing, E.P.T. and V.I.P. All authors have read and agreed to the published version of the manuscript.}

\funding{This research received no external funding.}

\acknowledgments{This paper has made use of: 
(a) the Hypatia Catalog Database, an online compilation of stellar abundance data as described in Hinkel et al. (2014, AJ, 148, 54), which was supported by NASA's Nexus for Exoplanet System Science (NExSS) research coordination network and the Vanderbilt Initiative in Data-Intensive Astrophysics (VIDA); 
(b) the NASA Exoplanet Archive, which is operated by the California Institute of Technology, under contract with the National Aeronautics and Space Administration under the Exoplanet Exploration Program;  
(c) the Planetary Systems Table, made available by the NASA Exoplanet Science Institute at IPAC, which is operated by the California Institute of Technology under contract with the National Aeronautics and Space Administration;  
and (d) the Exoplanet Orbit Database and the Exoplanet Data Explorer at  exoplanets.org.
}

\conflictsofinterest{The authors declare no conflict of interest.}

%
\appendixtitles{yes} 

\appendix
\appendixstart
\section{Brief Overview of Physics of ``Fission-Event''}\label{appa}

\subsection{Instability of Nuclear Matter: Nuclear-Fog}\label{appa1}
\unskip

Thermodynamical description of the states of the ``ordinary'' and nuclear matter 
is known to be similar in certain aspects, 
demonstrating the universality of fundamental laws of physics. 
For example, the normalized on critical temperature $T_c$ 
equations of state 
(pressure $P$ as a function of volume $V$) of a multi-body system of nucleons interacting via Skyrme potential  
is shown in Figure~\ref{Fig:A1}. 
\begin{figure}[H]
\includegraphics[width=0.50\columnwidth]{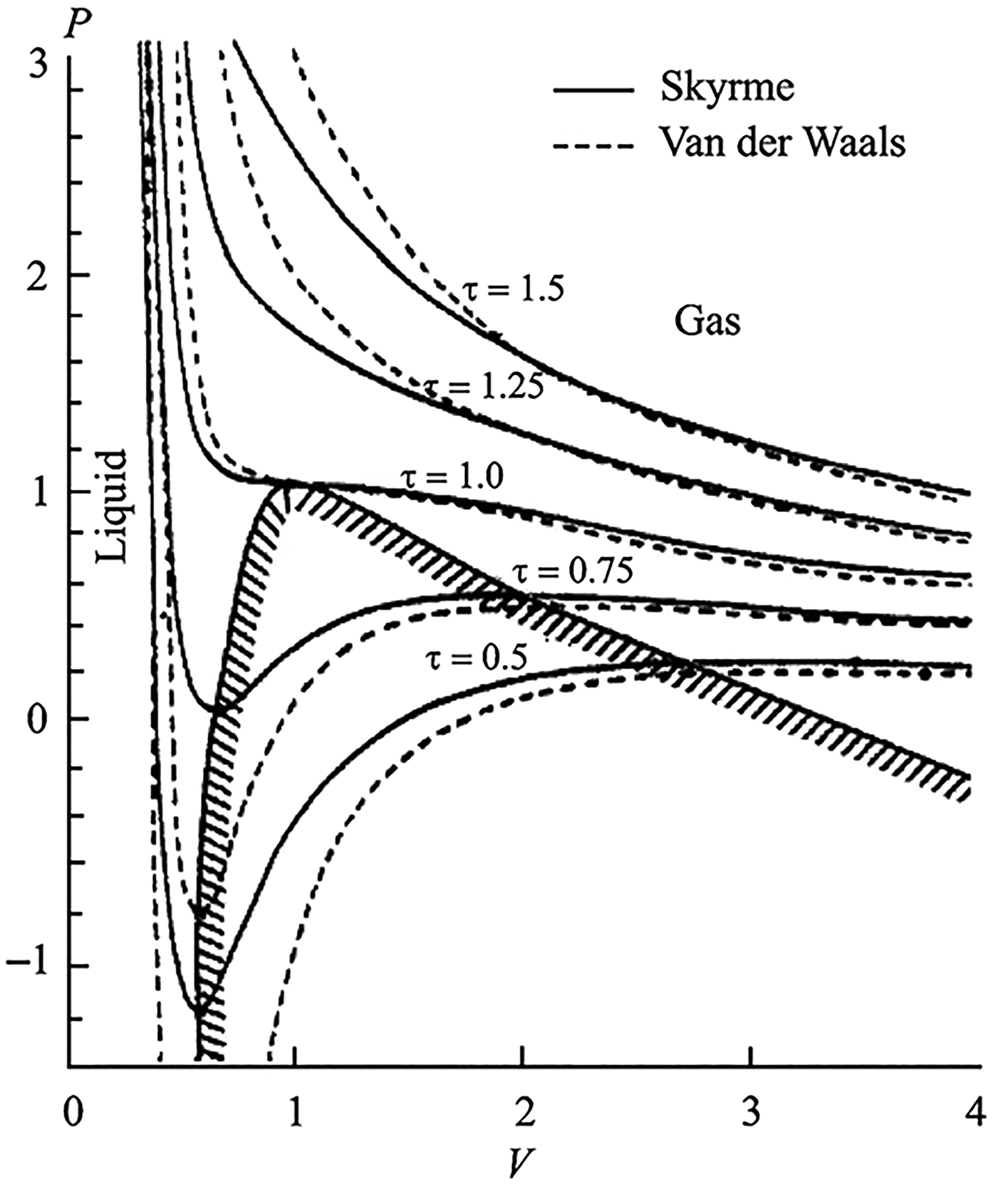} 
\caption
[Equations of state $P(V)$ for a nuclear system interacting through a Skyrme potential and a Van der Waals compressible liquid--gas system]
{
 From Ref. \cite{Jaqaman_1983}, equations of state $P(V)$ for a nuclear system interacting through a Skyrme potential and a Van der Waals compressible liquid--gas system (shown in relative units): qualitative similarity is apparent.  
}
\label{Fig:A1}
\end{figure}

Isotherms---$P(V)$ at constant temperature---corresponding to Skyrme effective interaction and finite temperature of Hartree--Fock theory (see \cite{Jaqaman_1983}) exhibit the maximum--minimum structure typical of the Van der Waals-like equation of state for a gas-liquid system of an ``ordinary'' (non-nuclear) matter.  
The very steep part of the isotherms (on the left side) corresponds to the liquid phase. 
The gas phase is presented by the right parts of the isotherms where pressure is changing smoothly with increasing volume. 
Between the gas and liquid zones lies the mixed zone where two phases can co-exist. 
For ordinary water, the mixed zone is the familiar water--fog. 
For nuclear matter, the mixture is the nuclear-fog---either liquid nuclear-droplets surrounded by gas of neutrons, or homogeneous neutron-liquid with neutron-gas bubbles. 
The crucial difference is that, if the nuclear-fog becomes ``sufficiently'' rarified, it ``explodes.''

The macroscopical behavior of the system within the zone where two phases can occur and co-exist is well-known. 
Two derivatives play important roles:  $(\partial P / \partial V )_T$ (at constant temperature $T$) and $(\partial P / \partial V )_S$ (at constant entropy $S$). 
In the spinodal zone (marked by the hatched line in Figure~\ref{Fig:A1}), 
where the isotherms correspond to the negative compressibility, i.e., $(\partial P / \partial V )_T > 0$,  
random density fluctuations evolve rapidly:
 the initially uniform system transforms into a mixture of two phases. Obviously, within the spinodal zone, the zone of collective instability (where the square of adiabatical speed $\sim (\partial P / \partial V )_S$ is negative) is inside the coexistence zone (where the square of isothermical speed $\sim (\partial P / \partial V )_T$ is negative).

The fact that nuclear matter may exist in the two-phase state has been known for a while \cite{Jaqaman_1983,Karnaukhov_2005}. 
Figure~\ref{Fig:A2} qualitatively depicts $T(\rho)$ phase diagram for nuclear matter. 
\begin{figure}[H]
\includegraphics[width=0.5\columnwidth]{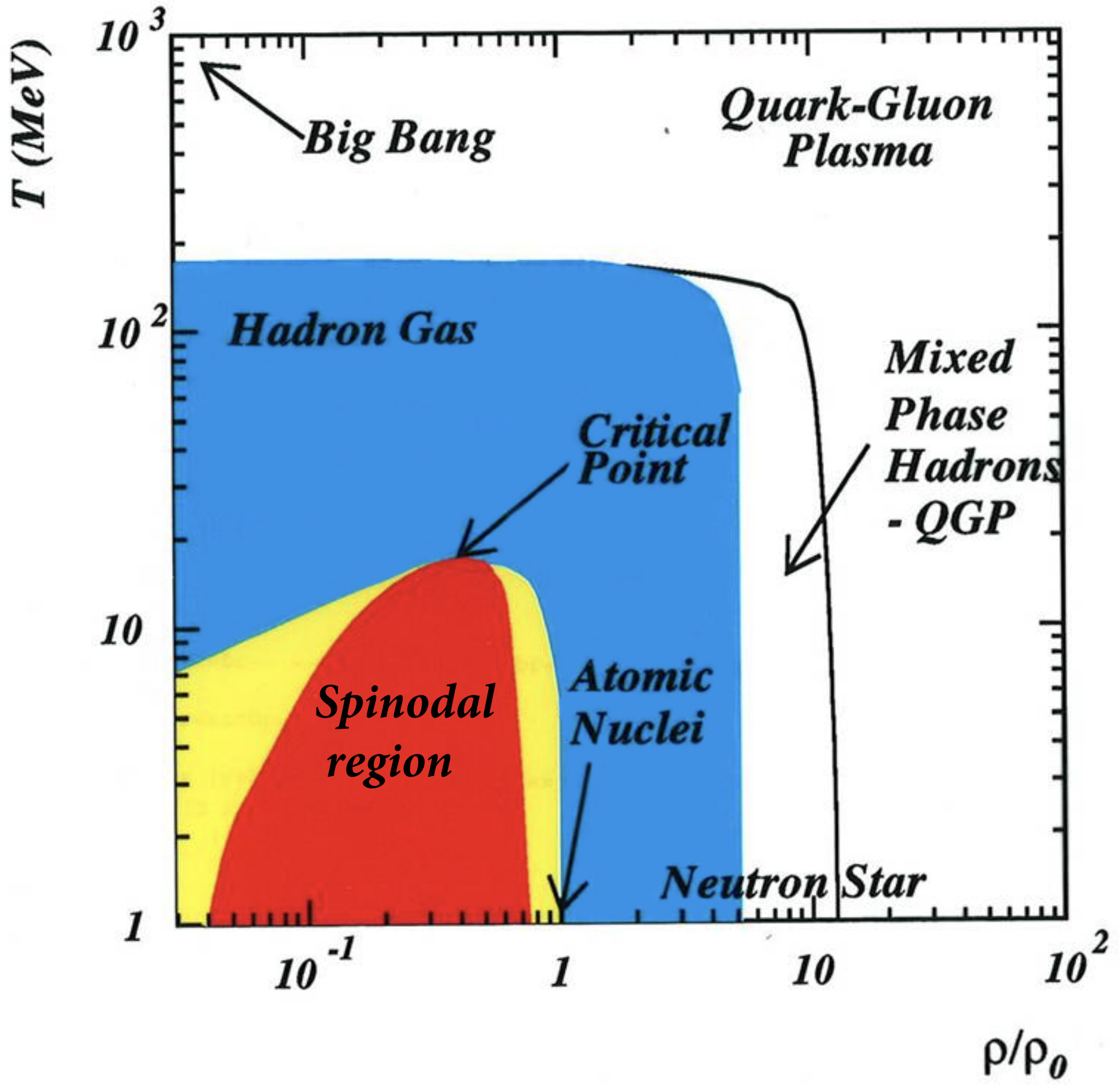} 
\caption
[Phase diagram $T(\rho)$ for nuclear matter]
{
 From Ref. \cite{Borderie_2019}.  Phase diagram $T(\rho)$ for nuclear matter  
(density $\rho$ is expressed in units of  $\rho_0 \equiv \rho_{nucleus} \simeq 2.85 \times 10^{14} \, g/cm^3 $,  
temperature $T$ is expressed in $MeV$ units, $1 \, MeV \simeq 10^{10} K$): 
the liquid--gas mixed phase region (yellow area, which ends up at the critical point) contains the spinodal region (red area). 
}
\label{Fig:A2} 
\end{figure}
High-energy nuclear experiments (in terrestrial conditions) have demonstrated that
the matter of a typical heavy-nuclei is characterized by 
the so-called 
critical parameters, such as temperature $T_c$ and density $\rho_c$.   
Over the years, experimental studies have provided a range of estimates for them:  
$\rho_c \simeq (0.1 \div 0.4)\rho_0$ and $T_c \sim 5 \div 18 \, MeV$ \cite{Karnaukhov_2006,Karnaukhov_2011}.  
Value $T_c = 17.5 \, Mev$ is commonly used. 
Notably, in laboratory conditions and experiments, parameters of nuclear targets are such that 
$T < T_c$ and $\rho_{nucl} \sim 2 \div 3 \rho_c$. 

\subsection{Structural Cohesion/Disintegration of Compact Super-Dense Stellar Fragment}\label{appa2}

As Figure~\ref{Fig:A2} indicates, the density of a neutron star (a super-giant-nucleus) is greater than the density of 
an ``ordinary'' nucleus---mainly due to gravitational effects.   
To compare, the mass of an ordinary neutron star is of the order of $M_{\odot} \sim 10^{30}$~(kg), while the mass of a nucleus with $A$ nucleons is $\sim$$10^{-27} \times A$~(kg).  

In a ground state (when $T/T_c = 0$, $\rho_0 / \rho_c > 1$), 
the nuclear matter is in the liquid state. 
However, as bombardment experiments with heavy nuclei (in terrestrial conditions) have revealed  
 (see Figure~\ref{Fig:A3},~left panel, from \cite{Karnaukhov_2006b}): 
«The van der Walls equation can be used with nuclear matter because of the similarity of the nucleon--nucleon force to the force between molecules in a classical gas ... 
In both cases, there exists a region in the PVT diagram corresponding to a mixture of liquid and gas phases. This region can contain unstable, homogeneous matter for short times. In a classical gas, this can be achieved by cooling through the critical point. In the nuclear case, this can be achieved by a sudden expansion of the liquid phase at a temperature well below the critical temperature. The separation of the homogeneous matter into a mixture of stable liquid and gas is called spinodal decomposition. One can imagine that a hot nucleus (at T = 7--10 MeV) expands due to thermal pressure and enters the unstable region. Due to density fluctuations, a homogeneous system is converted into a mixed phase consisting of droplets (IMF) and nuclear gas interspersed between the fragments. Thus, the final state of this transition is a nuclear fog ... 
Note that classical fog is unstable, and it condensates finally into bulk liquid. The charged nuclear fog is stable in this respect. However, it explodes due to Coulomb repulsion and is detected as multifragmentation». 
\cite{Karnaukhov_2006b}
(Here, IMF stands for ``intermediate mass fragments''.)  
Obviously, before entering the unstable $(T, \rho)$-region, 
the nucleus continues to remain structurally cohesive despite being excited (``hot'').  

A similar process occurs with a super-dense stellar fragment torn away by a massive black hole. 
(Perhaps a helpful metaphor for such cataclysm, even despite the differences in key ``holding'' forces, is a visual of a sizable ``ball'' of mercury  being 
tangentially hit with a baseball bat: the ball would ``spill'' as droplets of mercury, but not as a gas-like or dust-\mbox{like ``cloud''}.)

Even a ``small'' stellar-fragment (with the mass of $10^{-5} - 10^{-3} M_{\odot}$) is 
a giant-nucleus with $A \sim$$10^{52}-10^{54} $, which is 
50+ orders of magnitude more massive than an ordinary nucleus ($A \sim$$10^1-10^2$). 
Thus, strong gravity continues to play a substantial role in determining the fragment's 
density and structural 
cohesion. 
In comparison with the bombardment experiments on heavy nuclei (Figure~\ref{Fig:A3},~left panel), 
the stellar-fragment---born in a cataclysm---starts its life with much higher density and temperature than any nucleus considered in terrestrial laboratories. 
Nonetheless, the path-lines of the point representing the $(T, \rho)$-phase states 
are similar for both, the bombarded ordinary heavy nucleus and the stellar-fragment  
(Figure~\ref{Fig:A3},~left and right panels). 
For details, see Ref. \cite{Tito_2020}. 

\begin{figure}[H]  
\includegraphics[width=0.99\columnwidth]{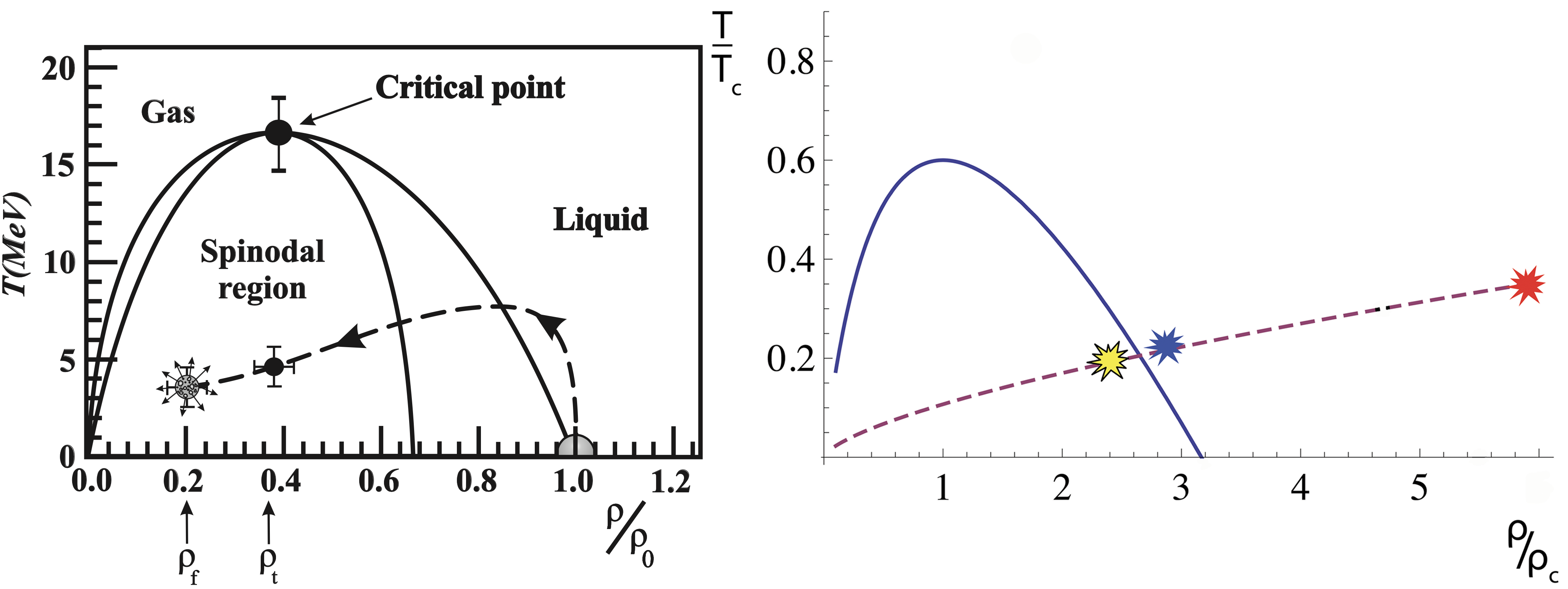}. 
      \caption{
{\bf Left Panel}:  From Ref. \cite{Karnaukhov_2006b}: Experimental data and analysis of fragmentation in $p(8.1GeV) + {\rm Au}$ collisions, and proposed spinodal region for the nuclear system.  The arrow line shows the path of the system from the starting point at $T=0$ and $\rho_0$ to the multi-scission point at $\rho_f$. 
Points  $\rho_t$ and $\rho_f$ denote the partition and freeze-out configurations.  
{\bf Right Panel:}  
Red, blue, and yellow stars represent stages of stellar-fragment evolution: 
red---''hot'' state after its birth, 
blue---''cool'' (stable) state near the spinodal zone boundary, 
yellow---''nuclear-fog'' (explosive) state. 
Adiabatic line $\theta = \theta_0 (z / z_0)^{2/3}$ is drawn to guide the eye.  
The actual path is more complex than \mbox{an adiabat}.
}
 \label{Fig:A3}
\end{figure}

One helpful reminder here to keep in mind is that a path in the $(T, \rho)$-plane represents changes in thermodynamical  phase states, not physical movements. 
In other words, ``entering the state of fog'' is not analogous to ``entering'' a sauna full of fog,  
but instead it is as analogous to ``watching'' how hot water in the sauna fills the room with fog (and later how the 
fog suddenly condenses on the walls when cool air comes in). 

\subsection{Multi-Fragmentation of Giant Nuclei}\label{appa3}

The process of multi-fragmentation of a ``giant nucleus may'' be schematically depicted as in Figure~\ref{Fig:A4}. 
(For transitions ``3'' and ``4''  in Figure~\ref{Fig:A4}, this schema is identical to the sketch for nuclei-bombardment experiments yielding multi-fragmentation of heavy \mbox{nuclei \cite{Karnaukhov_2012}}.)
\begin{figure}[H]
\includegraphics[width=0.70\columnwidth]{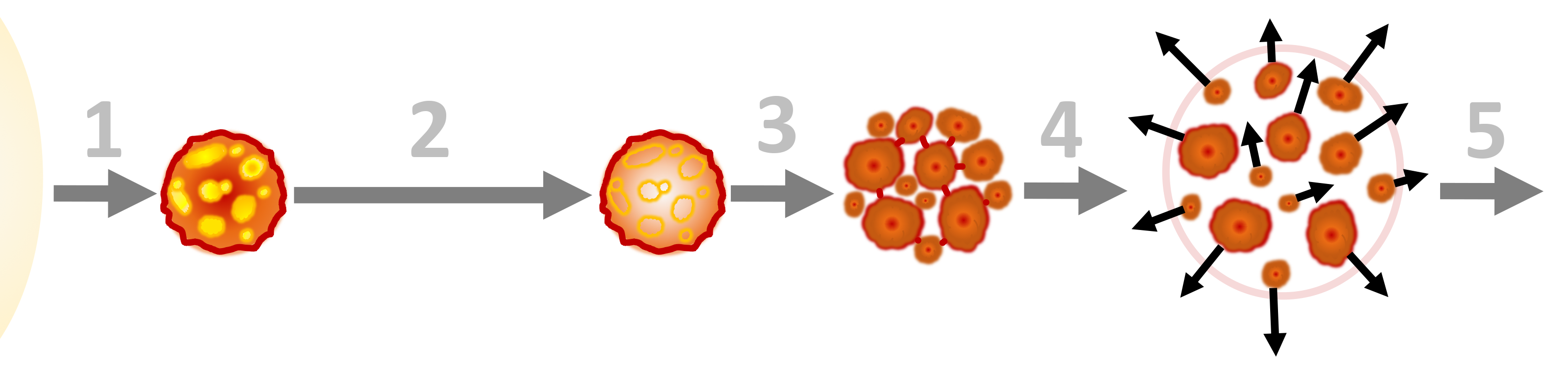}
\caption
{
Schema of multi-fragmentation of giant nuclei. 
Transition~``1'': disruption of a compact super-dense stellar-object (left) leading to ``birth'' of a stellar-fragment (right). 
Transition~``2'': cooling of the ``hot'' (left) stellar-fragment leading to ``cooler'' quasi-stable state (right).  
Transition~``3'': fluctuation/perturbation of the quasi-stable nuclear matter (left) and initiation of multi-fragmentation once unstable (right).  
Transition~``4'': explosive expansion (once $\beta$-decays become no longer Pauli-blocked).  
Transition~``5'': multi-fragmentation and fission of each hyper-nucleus (each droplet of nuclear fog), followed by chains of cascades (multi-fragmentation, fission, neutron-, $\beta$-, and $\gamma$-emissions; see Figure~\ref{Fig:1}, right panel, and Figure~\ref{Fig:2}) and dispersion of the material. Repetitions of steps 3--4--5 produce nuclei with lower and lower numbers of nucleons $A$, eventually reaching the valley-of-stability ($A\sim$$10^2-10^1$).
}
 \label{Fig:A4}
\end{figure}

Once born (during transition~``1''), the stellar-fragment starts with high initial temperature. 
The fragment is composed of pure neutrons (its charge $Z = 0$) and is held together, with high density, by nuclear forces and self-gravity. 
The fragment's cooling time may be long (depicted as transition ``2''  in Figure~\ref{Fig:A4}, and as the transition from the red-star state to the blue-star state in the right panel of Figure~\ref{Fig:A3}).  
(Generally speaking, the cooling process is complex,  
including neutrino-radiation,
and any estimates would require a sophisticated model, which is beyond the scope of this work.) 
Once ``sufficiently'' cool---which means that the point representing the fragment's $(T, \rho)$-phase-state crosses the boundary of nuclear-matter's phase-instability (spinodal) zone (Figure~\ref{Fig:A3}, right panel)---the evolution of the stellar-fragment proceeds analogously to the evolution of compound heavy-nuclei in the terrestrial experiments (Figure~\ref{Fig:A3}, left panel): the nuclear matter becomes ``nuclear fog'' and ``explodes''. 

The ``explosion'' of nuclear-fog is a process that is somewhat extended in time (still within nuclear-timescales).  
As mentioned, in the nuclear-fog state, the nuclear matter can exist as a mixture of two phases---either liquid droplets ($\equiv$ hyper-nuclei) surrounded by the gas of neutrons, or generally-homogeneous neutron-liquid with neutron-gas bubbles. 
Everything is charge-neutral. 
A random density fluctuation, or an externally-caused perturbation, triggers a rapid transformation of the initially uniform system into a mixture of these two phases. 
In such state, the matter can reach substantial further rarification, reducing density by a factor of $10^2$ or more due to collective hydrodynamic instability.  

Each nuclear-fog droplet (at first still charge-neutral, $Z=0$) is a giant nucleus composed purely of neutrons. 
As known, when a neutron is free or is within a neutron-oversaturated 
nucleus, the neutron is unstable and falls apart into a proton and an electron. 
This is the so-called $\beta$-decay. 
The process is accompanied by energy release.
Within an ensemble,  $\beta$-decay takes place {\em only} if density of 
   the ensemble (entirely or in some localized domain) 
is below the critical level $\rho_{drip}$  
(this is called Pauli-blocking).  
Once in some place the density falls below $\rho_{drip}$,  
neutrons start converting into protons (as electrons swiftly escape), and  
the initially non-charged (neutral) fog-droplet starts gaining some positive charge. 
 Nonetheless, overall and for a short while, the giant nucleus remains predominantly composed of neutrons (but its $Z/A$ ratio starts rising from zero to non-zero \mbox{and higher}). 

Thus,  
   once the quasi-stable (i.e., ``sufficiently'' cool) stellar-fragment is perturbed 
   (one possible perturbation mechanism, straight-line deceleration, is discussed in Ref. \cite{Tito_2020})
   and once in some domain the density falls 
below $\rho_{drip}$ (even if in 
a rather small physical domain within the gigantic nucleus),   
fragmentation of the supersaturated 
with neutrons 
hyper-nuclei 
    (nuclear-fog droplets) is no longer inhibited and starts cascading.   
These reactions, known to release substantial energy ($\sim$1 MeV per fission nucleon, as seen in
transuranium nuclei fission events), proceed effectively at the same moments as the $\beta$-decay reactions. 
Overall, the system undergoes 
cascades of simultaneous fragmentation/fission and 
neutron-, $\beta$-, and $\gamma$- emissions and captures. 
Everything happens very fast, with the time scales of the order of nuclear-timescales.

In short, as long as the giant nuclear-drop (stellar-fragment) is purely neutron ($Z = 0$), it is held together by nuclear forces and self-gravity  and remains charge-neutral. 
As soon as for some reason (for example, due to internal density perturbation when quasi-stable, followed by the mentioned $\beta$-decay) the ``drop'' gains some charge, it starts fragmenting/fissioning. 
In addition, so does each formed ``daughter“-fragment, which is also a hyper-nucleus, even if a smaller one than its ``parent''. 
(Generally speaking, the difference between multi-fragmentation and fission is that, 
for multi-fragmentation, 
the main decay mode of hot nuclei is a copious emission of intermediate mass fragments, which are heavier than $\alpha$-particles but lighter than fission fragments \cite{Karnaukhov_2006b}.)  

During the short time-interval of these fission-cascades, all sorts of emission and radiation processes go on at the same time and 
various super-heavy, heavy, light nuclei become created. 
Thus, the overall story 
of the traveling stellar-fragment 
 is not about ``a flying sack already full of various elements, crazy-heavy or not''. 
The elements (nuclei with non-zero $Z$) form {\em only} during the short-time of the ``encounter'',  
via fragmentation/fission cascades and accompanying decays/captures.

In summary, 
following the described process backwards (upward in $A$):  
the ``ordinary''  heavy nuclei (such as {\rm U}, for example, with $A \sim$$3 \times 10^2$) are produced by multi-fragmentation and fissioning (and various decays) of mega-nuclei (see \mbox{Figures~\ref{Fig:1} and~\ref{Fig:2}}).
The mega-nuclei  
are the products of multi-fragmentation/fission of their ``parents'', who are the ``daughters'' of their own multi-fragmenting/fissioning ``parents'', again and again, up to the hyper-nuclei which are the droplets of the nuclear-fog. 
The nuclear-fog is the state of the matter composing the stellar-fragment ($A \sim$$10^{50+}$) once the fragment is ``sufficiently'' cooled.  
The state of nuclear-fog lasts only for a short time once the density of the matter (even in a small localized internal domain) falls below some critical level.  
This can happen when the stellar-fragment cools down ``sufficiently'' or when the quasi-stable fragment runs into an ``obstacle'' which triggers internal density perturbations and subsequent development of instability. 
The stellar-fragment is a piece 
(initially non-charged and stable, until ``sufficiently cooled down'') 
torn away by a massive black hole from a compact super-dense stellar-object (i.e., composed of non-charged, purely neutron, matter). 
The cataclysm creating the stellar-fragment is denoted as transition~``1'' in Figure~\ref{Fig:A4}. 

\vspace{12pt}

\subsection{Origination of Stellar-Fragments}\label{appa4}

Transition~''1'' (in Figure~\ref{Fig:A4}) occurs in stellar cataclysms, 
such as in tidal disruptions of compact super-dense stellar objects. 
Generally speaking, a number of exotic compact stars have been hypothesized 
\cite{Shapiro_1983,Glendenning_2000}, 
such as:
quark stars---a hypothetical type of stars composed of quark matter, or strange matter;
electro-weak stars---a hypothetical type of extremely dense stars, in which the quarks are converted to leptons through the electro-weak interaction, but the gravitational collapse of the star is prevented by radiation pressure;
preon stars---a hypothetical type of stars composed of preon matter.   
Even dark energy stars  and  Planck stars  have been proposed.  
Other objects could exist billions of years ago. 
Neutron stars are the most commonly considered compact super-dense objects. 

In particular, 
stellar-fragments can be formed and  catapulted  
if a black hole tears a neutron star apart \cite{Rees_1990} without merging with it.  
Figure~\ref{Fig:A5}  illustrates the scenario.  

\begin{figure}[H]  
\includegraphics[width=0.99\columnwidth]{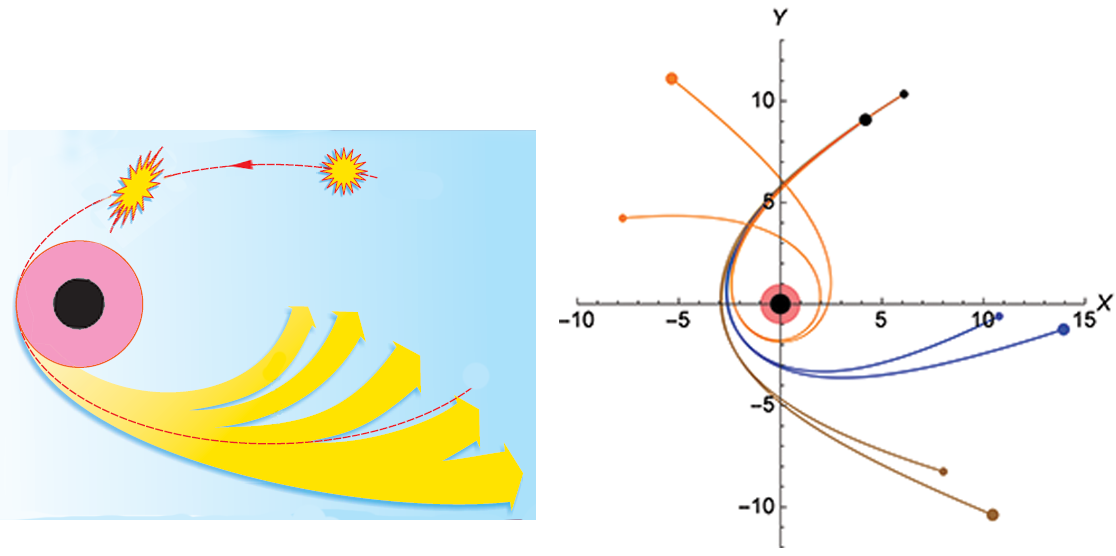}. 
      \caption{
{\bf Left Panel}: 
Adapted from \cite{Rees_1990}. 
 Conceptual illustration of tidal disruption of a fast-moving star by a massive rotating black hole 
(slicing along $\theta = 0$ surface reveals ``pink'' ergosphere and ``black'' black hole with its outer event-horizon) 
the star may be torn into pieces, some of which become captured, while others catapulted by the black hole.    
{\bf Right Panel:}  
Illustrative simulation of the process of  tidal destruction of a stellar body by a fast-rotating massive black hole: 
two fragments of the initially cohesive stellar body---depicted as the black dot near dimensionless coordinates (+6; +10)---follow different, bound or unbound, trajectories depending on the initial parameters of the system \cite{Tito_2018b}. 
}
 \label{Fig:A5}
\end{figure}

The supermassive black holes located at the centers of galaxies can ``crush'' neutron stars and catapult their ``fragments''. 
What is worth noting is that, because these central black holes have different masses, their powers as ``catapults'' should vary.  
Within the fission-events framework, the differences would produce different patterns of galactic enrichment.  
For galaxies whose central black holes are more massive, 
not only may more enrichment occur overall (because more neutron stars may be crushed), 
but also galactic 
peripheries may be more enriched 
(because the fragments may reach farther distances if the catapult is more powerful).  
This is consistent with that noted in \mbox{Section~\ref{s:3}} observations that 
 the ratios of {\rm [Y/Fe], [Ba/Fe], [La/Fe]}, and {\rm [Eu/Fe]} in the stars in the Milky Way dwarf spheroidal satellite galaxies (dSphs) and the stars in the Galaxy indicated that 
«{\rm [Y/Fe]} is significantly lower/offset in the dSph stars than in the Galaxy. This includes roughly half of the dSph stars, and suggests the $r$- and $s$-process enrichment of this element differs between the galaxies \dots 
This result suggests that  the site of $r$-processed {\rm Y} must differ from that of $r$-processed {\rm Ba, La}, and {\rm Eu}; is there a weak $r$-process site? In addition, the source that produces {\rm Y} in the metal-poor Galactic stars must be absent in the dSphs or it must have a different time lag relative to the {\rm Ba, La}, and {\rm Eu} enrichments  \dots 
[N]o population of stars in the Galaxy is representative of stars in the low mass dwarfs» 
\cite{Venn_2004}. 

In the fission-events framework, the enrichment patterns that are {\em similar} across galaxies may correspond to conventional production-sites, while the {\em differences} in the patterns across galaxies may point to the fission-events' production. 
Indeed, supernovae, AGB stars, and other conventional sites of nucleosynthesis, should work the same regardless of the properties of the black holes in the galaxy, while the frequency (and ``geography'') of fission-events may be correlated  with the ``power'' of black holes to crush and catapult fission-capable fragments. 

Another observation is that it would seem that any ``catapult'' is more likely to eject stellar-fragments {\em within the plane} of the galactic disk, so there should be enrichment asymmetry between the stars within the disk plane and the stars off-plane. 
Indeed, as previously mentioned, 
 «{\rm [Fe/H]}-rich group near the midplane is deficient in {\rm Mg, Si, S, Ca, Sc II, Cr II}, and {\rm Ni} as compared to stars farther from the plane» \cite{Hinkel_2014}, or, reversing the perspective, the midplane stars have excess {\rm Fe}.

\section{Likelihood of Fission-Event within Individual Planetary Systems} \label{appb}

\vspace{6pt}

As noted earlier, in the fission-events framework, any stellar-fragment
eventually explodes in the nucleogenetic cascades.  
For the general galactic enrichment, it does not matter where exactly such nucleogenesis occurs. 
However, for the search of ``impacted'' systems, particularly systems with planets, 
the question of encounter likelihood may arise.  
In Ref. \cite{Tito_2020}, we provided a discussion related to the solar system. 
To start, 
we noted 
that 
it is 
important to be clear what the term 'likelihood' is meant to describe.

\vspace{2pt}

The first kind of likelihood is `plausibility,' which inquires, in essence, whether the laws of physics permit the occurrence of the event in the first place. Understanding how a combination of various mechanisms can produce the event in question yields the conclusion that the event is {\em plausible}---in other words, {\em not impossible}, not forbidden by the laws \mbox{of physics.}

\vspace{2pt}

The second kind of likelihood is `statistical probability', which is about statistical odds of mental repetition of a {\em similar} event, not about whether the {\em first} (prior) event can happen. Questions about statistical probability always {\em imply} that the first event can or did happen. The concept of {\em statistical} probability of an event is connected with the concepts of the most {\em expected} outcome, the {\em frequency} of repeated events, and other related concepts. 
In the context of the solar system, 
 the ``statistical odds'' have nothing to do with the question of whether the  event proposed in our hypothesis 
could indeed have happened 4.6~Gyrs ago. Such an event would have been (was) the {\em first} event.  (and hence the only relevant inquiry is its plausibility.) 
In addition, we humans should be very happy that the odds of the {\em second} event happening in our solar system are low.

\vspace{2pt}

In addition,  when talking about probabilities, it is important to remember the difference between ``expectation'' and ``realization''. 
For an ``encounter'', and especially for a ``collision'', the often-used word ``target'' can mean two different things: 
the {\em intended-goal/specific aim for the path} (like the rope 
 for hanging 
which 
Clint Eastwood's 
hero was shooting from afar to release his co-conspirator in the movie «The Good, the Bad, and the Ugly») 
and the {\em accidental-result / random obstacle on the path} (like the hole that is left in a wall by a blind-man's accidental gunshot).  
Using these metaphors, we can say that  
the 
scenario 
of enrichment of the solar system 
is not about ``whether a bullet can hit the distant rope,'' 
but instead we note that ``the hole in the wall looks like it came from a bullet', so what kind of bullet must that have been and what might have happened.  
For accidental-results (obstacles), post-event, and statistical odds are irrelevant.  
Upon realization, P=1.
The same logic should apply when considering any exoplanetary system.

\vspace{2pt}

\textls[-25]{With respect to 
the probability numbers, 
the ``frequency of collisions'', \mbox{$\nu \equiv \tau^{-1} = n \langle \sigma V \rangle$}}, gives an indication about the \emph{chance}
of the occurrence of the event (collision) during some increment of time. Here, $n$ is a concentration of the obstacle population, 
$\sigma$ is interaction cross-section, and $V \times 1$ is the distance covered by the moving object over the unit of time. 
Properly speaking, 
expression $P = \nu \Delta t = \langle n \sigma V \rangle \Delta t$ is defined over the {\em large} number of possible realizations (where symbol $\langle ... \rangle$ denotes  statistical averaging, which is equivalent to ergodicity). A similar estimation is made, for example, for collisions between (microscopical) molecules of gas in a (macroscopical) container. 

\vspace{2pt}

For an event to occur on the outskirt of the galaxy, as far as the solar system is from the center  
(i.e., $V  \Delta t \sim$$3 \times 10^4$ light-years away), 
a traveling   
stellar-fragment 
with velocity $V \sim$$3 \times 10^{-3}$ of light-speed (i.e., $10^3$ km/sec) would 
need 
$10^7$ years---not too long of a time in comparison with the age of the universe ($\sim$$10^{10}$ years). 

\vspace{2pt}

Assuming $n \sim$$1^{-3}$ light-years$^{-3}$ (based on the average distance between stars in the central part of our galaxy $\sim$1 light-year) 
and  
$\sigma \sim$$(10^{-4})^2$ light-years$^2$ (which corresponds 
roughly to the area within Jupiter's orbit, implying that a ``collision'' may in fact ``perturb'' the 
stellar-fragment 
and the system, and thus end the journey), 
the frequency of such collisions is then $P \sim$$10^{-4} \ll 1$.  
(For more detailed considerations, see Ref. \cite{Binney_2008}.)
Long journeys, without any encounters along the way,  
provide stellar-fragments with enough time to cool down. 

\vspace{2pt}

An important note, however, is that the interaction 
cross-section for such encounters are not ``geometric''.  
The paths of stellar-fragments are always influenced by gravitational fields, 
so any massive ``centers''---large clusters of stars at far-distances, groups of stars at mid-distances, and individual massive stars nearby (or centers of mass in binary systems)---necessarily attract stellar-fragments.}


\end{paracol}
\reftitle{References}

\end{document}